\documentclass[aps,pre,preprintnumbers,showpacs,twocolumn,floatfix,superscriptaddress]{revtex4}
\preprint{LA-UR-14-21202}
\usepackage{amsmath,amssymb,dcolumn,eucal,graphicx,bm,color,hyperref}
\usepackage[caption=false,font=small]{subfig}
\newcommand{\be}{\begin{equation}}
\newcommand{\ee}{\end{equation}}

\begin{document}

\title{Successive phase transitions and kink solutions in 
$\phi^{8}$, $\phi^{10}$, and $\phi^{12}$ field theories}

\author{Avinash Khare}
\email{khare@iiserpune.ac.in}
\affiliation{Raja Ramanna Fellow, Indian Institute of Science Education
and Research, Pune 411021, India}

\author{Ivan C.\ Christov}
\email{christov@lanl.gov}
\affiliation{Theoretical Division and Center for Nonlinear Studies,
Los Alamos National Laboratory, Los Alamos, NM 87545, USA}

\author{Avadh Saxena}
\email{avadh@lanl.gov}
\affiliation{Theoretical Division and Center for Nonlinear Studies,
Los Alamos National Laboratory, Los Alamos, NM 87545, USA}

\date{\today}

\begin{abstract}
We obtain exact solutions for kinks in $\phi^{8}$,
$\phi^{10}$ and $\phi^{12}$ field theories with degenerate minima, which can describe a 
second-order phase transition followed by a first-order one, a succession of two 
first-order phase transitions and a second-order phase transition followed by two 
first-order phase transitions, respectively. Such phase transitions are known to 
occur in ferroelastic and ferroelectric crystals and in meson physics. In particular, 
we find that the higher-order field theories have kink solutions with algebraically-decaying 
tails and also asymmetric cases with mixed exponential-algebraic tail decay, 
unlike the lower-order $\phi^4$ and $\phi^6$ theories. Additionally, we construct 
distinct kinks with equal energies in all three field theories considered, and we
show the co-existence of up to three distinct kinks (for a $\phi^{12}$ potential with
six degenerate minima).  We also summarize phonon dispersion relations for these 
systems, showing that the higher-order field theories have specific cases in which only 
nonlinear phonons are allowed. For the $\phi^{10}$ field theory, which is a quasi-exactly 
solvable (QES) model akin to $\phi^6$, we are also able to obtain three analytical solutions
for the classical free energy as well as the probability distribution function in the 
thermodynamic limit.
\end{abstract}


\maketitle

\section{Introduction}

First- and second-order phase transitions are usually modeled by $\phi^{6}$ and
$\phi^{4}$ field theories, respectively \cite{makhan}.  An asymmetric double 
well in $\phi^4$ field theory can also describe first-order transitions \cite{sanati}.  
However, if one has to capture {\it all} symmetry-allowed phases in a 
low-dimensional phase transition \cite{toledano} or describe a succession of 
phase transitions, then one has to consider either 
multi-component field theories \cite{sonin} or higher-than-sixth-order 
single-component field theories \cite{lohe,gufan}. For example, it is well known 
\cite{toledano,gl} that while $\phi^{8}$ field theory can describe a second-order 
phase transition followed by a first-order phase transition, one has to go to 
$\phi^{10}$ field theory to describe a succession of two first-order phase
transitions.  Indeed, there are examples of crystals undergoing two 
successive (ferroelastic and ferroelectric) first-order phase transitions 
\cite{mroz}.  The $\phi^8$ field theory has also been used to model massless 
mesons with long-range interactions \cite{lohe} as well as isostructural 
phase transitions \cite{pavlov}.  Similarly, the $\phi^{10}$ field theory has 
been used in the study of crystallization of chiral proteins \cite{protein}.
Meanwhile, the $\phi^{12}$ field theory has been invoked  to describe the phenomenology of
phase transitions in highly piezoelectric perovskite materials \cite{vc,piezo}.

The study of kinks (also known as \emph{topological solitons} \cite{ms-ts})
and domain walls in classical and quantum field theories \cite{ps,Vach},
in theories of gravity and cosmology \cite{zeld,vm}
and even in the nonlinear field theories of fluid mechanics \cite{jordan} 
remains a topic of active research. Similarly, Ginzburg--Landau theories 
\cite{GinzburgLandau,tinkham} have been very 
successful in explaining superconducting, superfluid and many other transitions as 
well as in modeling topological defects (e.g., vortices and domain walls) in a variety 
of functional materials, through the inclusion of the gradient of the relevant order parameter in the free energy. 

In this context, solitary wave solutions of some special octic potentials have been presented 
before \cite{boya,yang}.  Similarly, generic properties of kink solutions of certain field theories with 
polynomial self-interaction have been studied previously \cite{casa,cooper,bazeia}.   
The purpose of this work is to provide the various kink solutions of the $\phi^{8}$, $\phi^{10}$  
and $\phi^{12}$ field theories with degenerate minima. In addition, we show that as in $\phi^{6}$ field theory 
(but unlike $\phi^{4}$ field theory), it is possible to obtain an exact expression for 
the classical free energy and probability distribution function (PDF) 
\cite{bk,bruce} at a 
given temperature in the thermodynamic limit of the $\phi^{10}$ field 
theory.  This is related to the fact that the Schr\"odinger equation with a 
$\phi^{4n}$ (e.g., $\phi^8$, $\phi^{12}$) potential is not analytically solvable, 
whereas with a $\phi^{4n+2}$ (e.g., $\phi^6$, $\phi^{10}$) potential
it is quasi-exactly solvable 
(QES)  \cite{leach}.

\section{$\phi^{8}$ Field Theory}

Throughout this paper, we refer to $\phi=\phi_e$ as an \emph{equilibrium} 
value if $V(\phi_e) = 0$, by \emph{degenerate extremum} we mean 
$\phi=\phi_e$ such that $V(\phi_e) = V'(\phi_e) = 0$,
all potentials are assumed symmetric, i.e., $V(-\phi) = V(\phi)$,
and we use Planck units ($m=c=\hbar=1$) to simplify the notation.

First, we will discuss the general picture, describing different possible phases
for a generic potential $V$,
then we will discuss the kink solutions in the various phases.

\subsection{The Various Phases}\label{sec:phi8_var_phas}

The $\phi^{8}$ potential (free energy) is given, generically, by
\be\label{1}
V(\phi) = \lambda^2(\phi^{8}-\alpha_6\phi^{6}+\alpha_4\phi^{4}
-\alpha_2\phi^{2}+\alpha_0),
\ee
where, 
%unless otherwise specified, $\alpha_{6,4,2} >0$ and, 
without loss of generality, the coefficient
of the $\phi^8$ term is set to $+1$ in units of $\lambda^2$.
The coefficients of $\phi^{6}$, $\phi^{4}$ and $\phi^{2}$ are, in general, arbitrary 
and there are eight different possibilities, depending on whether all three, 
two, one or none of the coefficients are positive. However, 
if one wants to consider a model describing a second-order transition 
followed by a first-order transition, then one must take $\alpha_{6,4,2}>0$ in 
(\ref{1}). Additionally, a particular choice of $\alpha_0$
ensures that the minimum value of the potential is zero, i.e., $\min_\phi V(\phi) = 0$.

\begin{figure*}%
\centering
\subfloat[]{\includegraphics[width=0.475\textwidth]{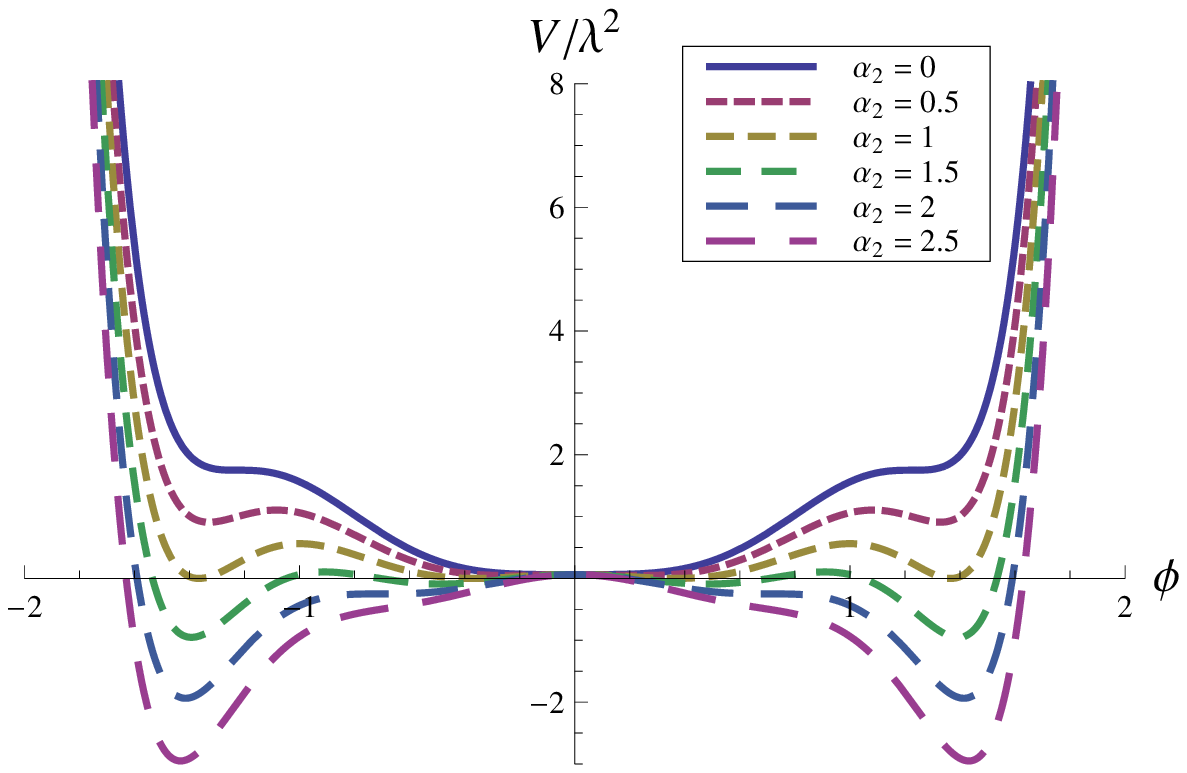}}\hfill
\subfloat[]{\includegraphics[width=0.475\textwidth]{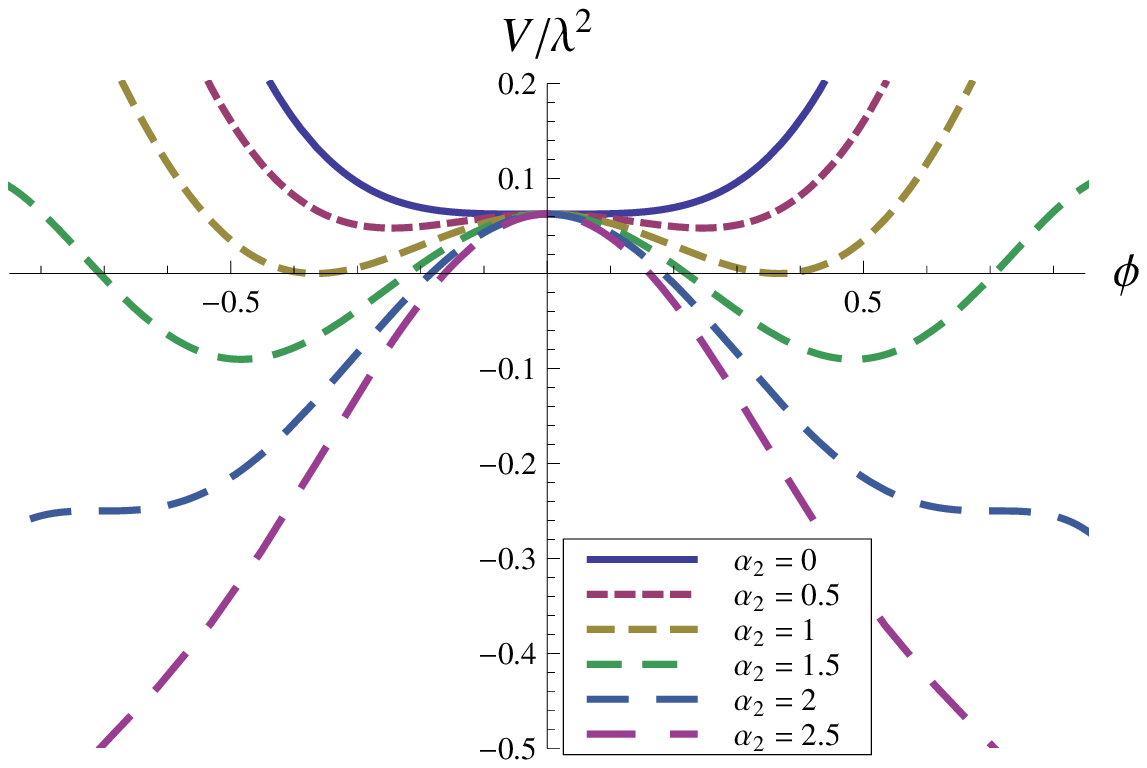}}
\caption{(Color online.) (a) Example potentials of the form (\ref{5}) for various illustrative values of the coefficient of the quadratic term, $\alpha_2$, showing the various phases and phase transitions in the $\phi^8$ theory. (b) Zoom-in of (a) near the origin.}
\label{fig:phi8_fig1}
\end{figure*}

While the potential is determined by three parameters ($\alpha_{6,4,2}$), one
can show, by using scaling arguments, that only two of them are  
independent. In a Landau-type theory, the coefficients $\alpha_{6,4,2}$ have
some dependence on the thermodynamic temperature $T$.
Thus, at the first-order phase transition point $T=T^{I}_c$,
$V$ has four degenerate minima, and the coefficients $\alpha_{6}$, 
$\alpha_{4}$ and $\alpha_{2} = \alpha_{2}^c$ are related by
\be\label{2}
\alpha_4 = \frac{\alpha_6^2}{4}+\frac{2\alpha_2^{c}}{\alpha_6}.
\ee
In particular, if the four degenerate minima are at $\phi=\pm a, \pm b$, then
the potential at $T=T^{I}_c$ has the factorized form
\be\label{4}
V(\phi)= \lambda^2 (\phi^2-a^2)^2(\phi^2-b^2)^2. 
\ee
Without loss of generality, we choose $b > a$ throughout this paper,
unless otherwise specified.
On comparing (\ref{1}) and (\ref{4}), and enforcing $\min_\phi V(\phi)=0$, 
it is clear that the relationship between $\alpha_{6,4,2,0}$ and $a,b$ is given by
\be\label{3}
\begin{aligned}
\alpha_6 &= 2(b^2+a^2),\\ 
\alpha_4 &= (b^4+a^4+4a^2b^2),\\
\alpha_2 &= \alpha_2^{c} = 2a^2b^2(b^2+a^2),\\
\alpha_0 &= a^4b^4.
\end{aligned}
\ee
Clearly, $\alpha_{6,4,2,0}>0$.
In this case, one can also show that the potential has
maxima at $\phi=0, \pm \sqrt{(b^2+a^2)/2}$. 
From (\ref{3}), we also find that $\alpha_4 /\alpha_6^2$ is 
constrained to satisfy the inequality
\be\label{4a}
\frac{1}{4} < \frac{\alpha_4}{\alpha_6^2} < \frac{3}{8}.
\ee

Now, what happens as $T$ is slowly increased from $T^{I}_c$? It 
is easily shown that, in this model, keeping $\alpha_6,\alpha_4$ fixed and 
decreasing $\alpha_2$ (from its value $\alpha_2^{c}$ at $T=T^{I}_c$), 
$T$ goes above $T^{I}_c$. As soon as $T$ is slightly greater than $T^{I}_c$,
the potential has two degenerate absolute minima at $\phi=\pm \hat{a}$,
where $0 < \hat{a} < a$. Furthermore, 
there are now two local minima at $\phi=\pm \hat{b}$, 
where $0<\hat{b} < b$, and there are three maxima including one at $\phi=0$. 
As the temperature is further increased, the degenerate absolute minima
at $\phi=\pm \hat{a}$ persist until the onset of the second-order transition at 
$T=T_c^{II}$, which corresponds to $\alpha_2=0$. Beyond this point
($T>T_c^{II}$), the absolute minimum is now at $\phi=0$, not at
$\phi=\pm \hat{a}$. Meanwhile, it can be shown that, as long as 
$1/4 < \alpha_4/\alpha_6^2 < 9/32$, there are local minima at 
$\phi=\pm \hat{b}$ even at $T=T_c^{II}$ (i.e., $\alpha_2=0$);
if $\alpha_4/\alpha_6^2 =9/32$, then for $\alpha_2=0$, there are inflection points  
at $\phi=\pm \hat{b}$. However, if $9/32 < \alpha_4/\alpha_6^2 < 3/8$, 
then, as the temperature is slowly increased from $T_c^{I}$, the 
local minima at $\phi=\pm \hat{b}$ disappear even before the second
order transition point $T=T_c^{II}$ (i.e., $\alpha_2=0$) is reached.

Let us now discuss what happens as temperature is lowered from 
$T_c^{I}$, i.e., $\alpha_2$ is increased from its value $\alpha_2^{c}$ 
at $T_c^{I}$ (while keeping $\alpha_6$, $\alpha_4$ fixed). 
As soon as $T$ is slightly less than $T_c^{I}$, 
the potential has degenerate absolute minima at $\phi=\pm \hat{b}$, 
where $\hat{b}>b$, while there are degenerate local minima at $\phi=\pm \hat{a}$,
where $\hat{a}>a$, and three maxima 
including one at $\phi=0$. Finally, beyond a critical point, the 
local minima at $\phi = \pm \hat{a}$ disappear, and the potential only has absolute
minima at $\phi = \pm \hat{b}$ and a maximum at $\phi=0$. This picture
persists no matter how much further the temperature is lowered  (i.e.,
$\alpha_2$ is increased). For example, it is easily shown that at
$\alpha_2=2\alpha_2^{c}$ (for given $\alpha_6$, $\alpha_4$), 
the potential (\ref{1}) can be written as
\be\label{4b}
V(\phi)=\lambda^2 [\phi^2-(b^2+a^2)]^2[\phi^4+2a^2b^2].
\ee
Hence, at $\alpha_2=2\alpha_2^{c}$, $V$ has absolute
minima at $\phi=\pm\sqrt{b^2+a^2}$, a maximum at $\phi=0$ and {\it no} local
minima as long as $(b^2+a^2)^2 < 16a^2b^2$. Using (\ref{3}), it
follows that the
local minima at $\phi= \pm a$ disappear for some value of $\alpha_2<
2\alpha_2^{c}$ if $9/32 < \alpha_4/\alpha_6^2 <
3/8$. On the other hand, if 
$1/4 < \alpha_4/\alpha_6^2 < 9/32$, 
then $V$ has local minima at $\phi=\pm
\hat{a}$ with $a<\hat{a}$, while for $\alpha_4/\alpha_6^2=9/32$, $V$ has
inflection points at $\phi=\pm\sqrt{(b^2+a^2)/2}$. 

As an illustration, consider the potential 
\be\label{5}
V(\phi)=\lambda^2 [\phi^{8} - 4\phi^6 + (9/2) \phi^4 - \alpha_2 \phi^2 +(1/16)]
\ee
for various values of the parameter $\alpha_2$. For
$\alpha_2=\alpha_2^c=1$ this potential has four degenerate minima
with $b^2+a^2=2$ and $b^2-a^2=\sqrt{3}$ [see (\ref{4}) and (\ref{3})], hence 
this case corresponds to the first-order phase transition at $T=T_c^{I}$. 
Furthermore, for the potential in (\ref{5}), $\alpha_4=(9/32)\alpha_6^2$, hence at
$T=T_c^{II}$ (i.e., $\alpha_2=0$), $V$ has an absolute minimum at $\phi=0$ and
inflection points at $\phi=\pm\sqrt{(3/4)(b^2+a^2)}=\pm\sqrt{3/2}$. Similarly, for
$\alpha_2 =2\alpha_2^{c}=2$, there are absolute minima at $\phi=\pm\sqrt{2}$,
a maximum at $\phi=0$ and points of inflection at $\phi=\pm\sqrt{(b^2+a^2)/2}=\pm1$.
Thus, for $0 < \alpha_2 < 1$, the potential (\ref{5}) has absolute minima
at $\phi = \pm \hat{a}$, local minima at $\phi= \pm \hat{b}$ and
three maxima, including one at $\phi=0$. Similarly, for $1 < \alpha_2 < 2$, 
the potential has absolute minima
at $\phi = \pm \hat{b}$, local minima at $\phi= \pm \hat{a}$ and
three maxima, including one at $\phi=0$. For $\alpha_2 <0$, the potential has a single
minimum at $\phi=0$, while for
$\alpha_2 >2$, the potential has degenerate minima at $\phi = \pm b$, a 
maximum at $\phi=0$ and no local minima. In Fig.~\ref{fig:phi8_fig1}, we show plots
of  the example potential \eqref{5} 
for $\alpha_2=0,0.5,1,1.5,2,2.5$ (in units of $\lambda^2$) to illustrate its structure.

\subsection{Four Degenerate Minima}
\subsubsection{$T=T_c^{I}$}
  
At the first-order phase transition, i.e., $T=T_c^{I}$, the potential 
can always be written in the form (\ref{4})
with $a,b$ and $\alpha_{6,4,2}$ being related by (\ref{2}).  
Since there are four degenerate minima, we
expect two different kinds of kinks, one connecting $a$ to $b$ (or, equivalently, $-b$ to $-a$) 
and another connecting $-a$ to $+a$, as $x$ goes from $-\infty$ to $+\infty$. 
In general, these kinks have different energies. Notice that if a kink goes from $a$ to $b$, 
as $x$ goes from $-\infty$ to $+\infty$, then the kink's energy is given by
\be\label{4x}
E_k = \int_{-\infty}^{+\infty} dx \left[\frac{1}{2}\left(\frac{d\phi}{dx}\right)^2+V(\phi)\right]
=\int_{a}^{b} d\phi\, \sqrt{2V(\phi)},
\ee
where the last equality follows from the first integral of the equation
of motion, i.e., $d\phi/dx = \sqrt{2V(\phi)}$ \cite{lohe, BishopPhysD}; again,
without loss of generality, it is assumed that $\min_\phi V(\phi) = 0$, i.e., 
$V(\phi) \ge 0$ for all $\phi$, which is always true when kink solutions exist.

From the first integral of the equation of motion \cite{footnote1},
the shape of the kink can be found by quadrature:
\be\label{4.1}
\sqrt{2} \lambda x = \int \frac{d\phi}
{\sqrt{(a^2-\phi^2)^2(b^2-\phi^2)^2}}.
\ee
As mentioned above, there are two kinds of kinks 
to be considered, which leads to two possible choices in the branch cut of the square 
root in (\ref{4.1}). Let us consider the two cases separately.

\paragraph{Kink connecting $-a$ to $+a$}
In this case, $|\phi| < a$ and $b>a$ by convention, hence (\ref{4.1}) becomes
\be\label{4.1a}
\sqrt{2} \lambda x = \int \frac{d\phi}
{(a^2-\phi^2)(b^2-\phi^2)}.
\ee
The integral is evaluated using partial fractions to obtain the implicit solution,
which was also found by Lohe \cite[Eq.\ (63)]{lohe}:
\be\label{4.2}
e^{\mu x} = \left (\frac{a+\phi}{a-\phi} \right )
\left (\frac{b-\phi}{b+\phi} \right )^{a/b},
\ee
where $\mu = 2\sqrt{2}\lambda a (b^2-a^2)$. 
The approach to the asymptotes
at $\phi = \pm a$ can be shown to be exponential from (\ref{4.2}):
\be
\phi(x) \simeq 
\begin{cases}
-a + 2a \left(\frac{b-a}{b+a}\right)^{a/b} e^{\mu x},\quad &x\to -\infty,\\[3mm]
+a - 2a \left(\frac{b-a}{b+a}\right)^{a/b} e^{-\mu x},\quad &x\to +\infty,
\end{cases}
\ee
from which it follows that this kink is symmetric.
The corresponding kink energy is obtained using (\ref{4x}):
\be\label{4x1}
E_{k}^{(1)} = \frac{4\sqrt{2}}{15} \lambda a^3(5b^2-a^2).
\ee

\begin{figure*}
\centerline{\includegraphics[width=0.8\textwidth]{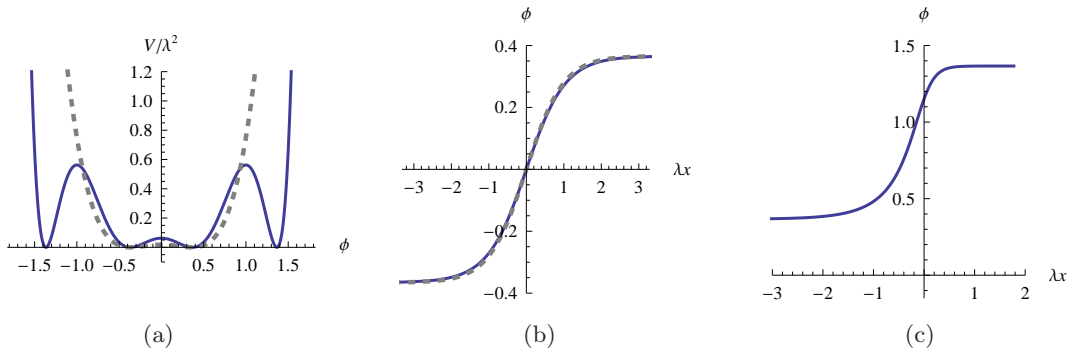}}
\hspace{3.6cm}(a)\hfill(b)\hfill(c)\hspace{3.8cm}
\caption{(Color online.)  $\phi^8$ field theory at the first-order phase transition, $T=T_c^I$. (a) Example $\phi^8$ potential with four degenerate minima \eqref{4} and a representative $\phi^4$ potential $V(\phi) = \lambda^2(\phi^2-a^2)^2$ superimposed as the dotted curve. (b) Kink solution \eqref{4.2} connecting $-a$ to $+a$ and the generic $\phi^4$ kink $\phi(x) = \frac{1}{2}(\phi_{+\infty} + \phi_{-\infty}) + \frac{1}{2}(\phi_{+\infty} - \phi_{-\infty}) \tanh (\lambda x) = a \tanh (\lambda x)$ for $\phi_{\pm\infty} = \pm a$ superimposed as the dotted curve. (c) Kink solution \eqref{4.2b} connecting $a$ to $b$. In all panels, $a=(-1+\sqrt{3})/2$ and $b=(1+\sqrt{3})/2$.}
\label{fig:phi8_TCI_kinks}
\end{figure*}

\paragraph{Kink connecting $a$ to $b$ (or $-b$ to $-a$)}
\label{sec:phi8-tcI-a}
In this case, $a < \phi < b$ and $b > a$ by convention, hence (\ref{4.1}) takes the form
\be\label{4.1b}
\sqrt{2} \lambda x = \int \frac{d\phi}
{(\phi^2-a^2)(b^2-\phi^2)}.
\ee
The integral is again evaluated using partial fractions to obtain the implicit solution
\be\label{4.2b}
e^{\mu x} = \left (\frac{\phi-a}{\phi+a} \right )
\left (\frac{b+\phi}{b-\phi} \right )^{a/b},
\ee
where $\mu = 2\sqrt{2}\lambda a (b^2-a^2)$ as before. The approach to the asymptotes
at $\phi = a, b$ can be shown to be exponential from (\ref{4.2b}):
\be
\label{4.2asymp}
\phi(x) \simeq 
\begin{cases}
a + 2a\displaystyle \left(\frac{b-a}{b+a}\right)^{a/b} e^{\mu x},\quad &x\to -\infty,\\[3mm]
b - 2b\displaystyle \left(\frac{b-a}{b+a}\right)^{b/a} e^{-\mu x b/a},\quad &x\to +\infty.
\end{cases}
\ee
%Analogous expressions can be derived for the asymptotes at $\phi = -b, -a$.
Note, however, that the rate at which $\phi$ asymptotes to $a$ is given by $\mu$,
while the rate at which $\phi$ asymptotes to $b$ is given by $\mu b/a$,
hence this kink is {asymmetric}.
The kink's energy is
\be\label{4.2d}
E^{(2)}_{k} = \frac{2\sqrt{2} }{15}\lambda (b-a)^3 (b^2+3ab+a^2).
\ee

Comparing the energies of the two kink solutions [(\ref{4x1}) and (\ref{4.2d})], we find that 
$E_{k}^{(1)} \gtreqqless E_{k}^{(2)}$ if $b/a \lesseqqgtr 2/(3-\sqrt{5})$.
In particular, for $b/a =  2/(3-\sqrt{5})$, the two kinks have equal energies.
It would be of interest to study the interaction between two
kinks of the same type as well as two kinks of different type
in the case when their energies are equal.

As an illustration, consider the potential (\ref{4}) with $a^2+b^2=2$ and $a^2 b^2=1/4$
so that $\alpha_2^{c}=1$. This leads to eight possible pairs $(a,b)$
with four of them satisfying $b^2>a^2$. Without loss of generality, we also take 
$a>0$ and $b>0$, hence $a=(-1+\sqrt{3})/2$ and $b=(1+\sqrt{3})/2$. 
Figure~\ref{fig:phi8_TCI_kinks} shows the potential (\ref{4}) and the two kink solutions
(\ref{4.2}) and (\ref{4.2b}). The kink solution from (\ref{4.2b}) is clearly
asymmetric, consistent with the asymptotic behaviors given in (\ref{4.2asymp}).

\subsubsection{$T_c^{I} < T < T_c^{II}$}

For temperatures above the first-order phase transition,
the potential \eqref{1} can be written as
\be\label{51}
V(\phi)=\lambda^2 (\phi^2-\hat{a}^2)^2 [\phi^4 -d\phi^2+e],
\quad d^2 < 4e,\quad \hat{a} < a,
\ee
and there exists a kink solution connecting the two degenerate minima
at $\phi=\pm\hat{a}$, as $x$ goes from $-\infty$ to $+\infty$.
As an illustration, consider the potential (\ref{5}) with 
$\alpha_2=121/128$. In this case, \eqref{51} takes the form
\be\label{6}
%\begin{aligned}
V(\phi) = \lambda^2 [\phi^2-(1/8)]^2[\phi^4-(15/4)\phi^2+(227/64)].\\
%&=\lambda^2 [\phi^{8} - 4\phi^6 +4.5 \phi^4 - (121/128)\phi^2 +(227/4096)].
%\end{aligned}
\ee
 
\subsubsection{$T<T_c^{I}$}
\label{sec:phi8-t_less_TcI}

For temperatures below the first-order phase transition,
the potential \eqref{1} can be written as
\be\label{52}
V(\phi)=\lambda^2 (\phi^2-\hat{b}^2)^2 [\phi^4 -d\phi^2+e],
\quad d^2 < 4e,\quad b < \hat{b},
\ee
and there exists a kink solution connecting the two degenerate minima
at $\phi=\pm\hat{b}$, as $x$ goes from $-\infty$ to $+\infty$.
As an illustration, consider the potential with $\alpha_2=135/128$. In this
case, \eqref{52} takes the form
\be\label{7}
%\begin{aligned}
V(\phi) = \lambda^2 [\phi^2-(15/8)]^2[\phi^4-(1/4)\phi^2+(3/64)].\\
%&=\lambda^2 [\phi^{8} - 4\phi^6 +4.5 \phi^4 - (135/128)\phi^2 +(675/4096)].
%\end{aligned}
\ee

\begin{figure}[h]
\centerline{\includegraphics[width=0.5\textwidth]{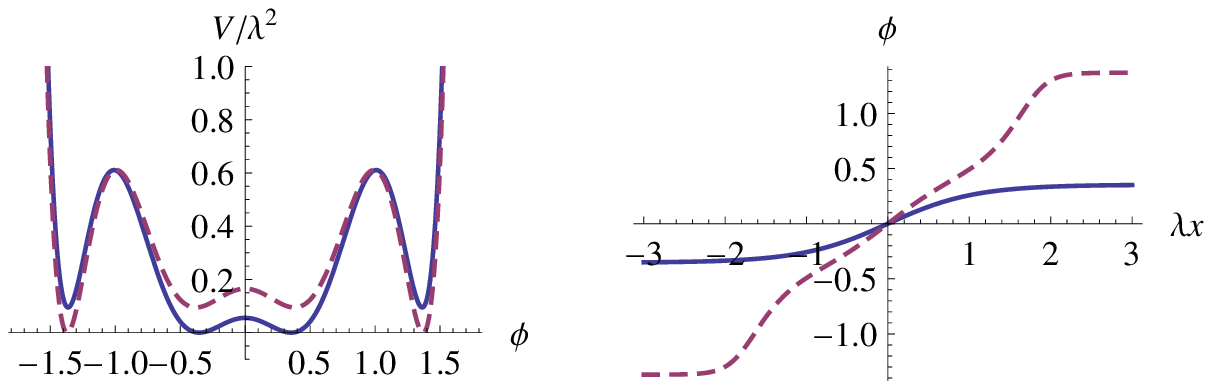}}
\hspace{1.65cm}(a)\hfill(b)\hspace{2cm}
\caption{(Color online.) Away from the first-order phase transition, $T \ne T_c^I$, in $\phi^8$ field theory. (a) The potentials (\ref{6}) (solid, $T>T_c^I$) and (\ref{7}) (dashed, $T<T_c^I$). (b) The corresponding kinks computed by solving the equation of motion $d\phi/dx = \sqrt{2V(\phi)}$ numerically subject to the symmetry condition $\phi(0) = 0$.}
\label{fig:phi8_nonTCI_kinks}
\end{figure}

These kink solutions for $T\gtrless T_c^I$ are illustrated in 
Fig.~\ref{fig:phi8_nonTCI_kinks}. Notice that for the case 
$T<T_c^I$ (dashed kink in right panel of Fig.~\ref{fig:phi8_nonTCI_kinks}), 
the kink ``feels'' the influence of the two local minima at 
$\phi = \pm \frac{1}{4}\sqrt{3(3-\sqrt{5})}\approx \pm 0.378$, similarly to kinks in 
certain cases of $\phi^6$ field theory \cite{sanati99}.

\subsection{Three Degenerate Minima}

A $\phi^8$ potential with three degenerate minima can have two possible forms.
In each case, there exist two kink solutions, only one of which is distinct due to
the symmetry of the potential.

\subsubsection{Case I: $\alpha_2 = 0$}
Let
\be\label{7.1}
V(\phi) = \lambda^2 \phi^4 (\phi^2-a^2)^2,
\ee
which has degenerate minima at $\phi=0, \pm a$.  In this case,
$\alpha_{6,4} > 0$ while $\alpha_{2,0} =0$. The 
corresponding kink solution (connecting $0$ to $a$ or $-a$ to $0$, as $x$ goes from 
$-\infty$ to $+\infty$), which was also obtained by Lohe \cite[Eq.\ (67)]{lohe},
 is given implicitly by
\be\label{7.2}
\mu x = -\frac{2a}{\phi}+\ln \left (\frac{a+\phi}{a-\phi} \right ),
\ee
where $\mu = 2\sqrt{2} \lambda a^3$. [It should be noted that
there is a typographical error in \cite[Eq.~(67)]{lohe} that is evident upon
comparison with \eqref{7.2}.]
From (\ref{7.2}), the approach to the 
asymptotes at $\phi = 0, a$ can be shown to be
\be\label{7.2asymp}
\phi(x) \simeq 
\begin{cases}
-\displaystyle \frac{2a}{\mu x},\quad &x\to -\infty,\\[3mm]
a - \displaystyle\frac{2 a}{e^2} e^{- \mu x },\quad &x\to +\infty.
\end{cases}
\ee
%Analogous expressions can be derived for the asymptotes at $\phi = -a, 0$.
Note that this kink is asymmetric because the asymptotics as $x\to\pm\infty$ 
differ, specifically the kink decays as $1/x$ as $x\to-\infty$, while it approaches 
$\phi = a$ exponentially.
The corresponding kink energy is
\be\label{7.2a}
E_k = \frac{2\sqrt{2}}{15}\lambda a^5.
\ee

\subsubsection{Case II: $\alpha_2 < 0$}
Let
\be\label{7.3}
V(\phi) = \lambda^2 \phi^2 (\phi^2-a^2)^2 (\phi^2+b^2),
\ee
which has degenerate minima at $\phi=0, \pm a$. Note that, in this case,
\be
\begin{aligned}
\alpha_{6} &= b^2 - 2a^2,\\
\alpha_{4} &= a^2 (a^2 - 2b^2),\\
\alpha_{2} &= -a^4b^2,\\
\alpha_{0} &= 0.
\end{aligned}
\ee
It can be shown that $\alpha_{6} > 0$ as long as $b>\sqrt{2}a$, while $\alpha_{4,2}  < 0$. 
The corresponding kink solution (connecting 
$0$ to $a$ or $-a$ to $0$ as $x$ goes from $-\infty$ to $+\infty$) is 
given implicitly by
\begin{multline}\label{7.4}
e^{\mu x} = \bigg (\frac{\sqrt{b^2+\phi^2}-b}
{\sqrt{b^2+\phi^2}+b} \bigg )^{\sqrt{b^2+a^2}/b}\\
\times \bigg (\frac{\sqrt{b^2+a^2}+\sqrt{b^2+\phi^2}}
{\sqrt{b^2+a^2}-\sqrt{b^2+\phi^2}} \bigg ),
\end{multline}
where $\mu = 2\sqrt{2} \lambda a^2 \sqrt{b^2+a^2}$.
From (\ref{7.4}), the approach to the 
asymptotes at $\phi = 0, a$ can be shown to be
%\begin{widetext}
\be\label{7.4asymp}
\phi(x) \simeq 
\begin{cases}
2b \left[ 1 + \frac{2b}{a^2}(b+\sigma) \right]^{-b/(2\sigma)} e^{\mu x b/(2\sigma)},\quad &x\to -\infty,\\[3mm]
a - \frac{2\sigma}{a}\left[1+\frac{2b}{a^2}(b-\sigma) \right]^{\sigma/b} e^{- \mu x },\quad &x\to +\infty,
\end{cases}
\ee
%\end{widetext}
where $\sigma =\sqrt{b^2+a^2}$.
%Analogous expressions can be derived for the asymptotes at $\phi = -a, 0$.
Note the differing rates $\mu b/(2\sigma)$ and $\mu$ at which 
the asymptotes at $x\to\mp\infty$, respectively,  are approached, 
hence this kink is asymmetric in general.
The corresponding kink energy is
\be\label{7.4b}
E_k = \frac{\sqrt{2}}{15} \lambda
\left[2(b^2+a^2)^{5/2} -b^3(2b^2+5a^2) \right].
\ee

\begin{figure}[h]
\centerline{\includegraphics[width=0.5\textwidth]{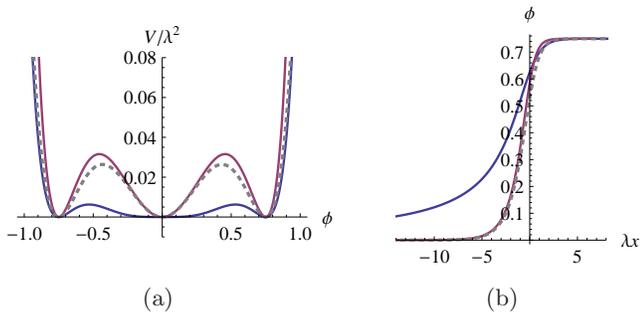}}
\hspace{1.65cm}(a)\hfill(b)\hspace{2cm}
\caption{(Color online.) $\phi^8$ field theory with three degenerate minima. (a) The potentials (\ref{7.1}) (bottom curve, blue online), (\ref{7.3}) (top curve, red online), and a representative  $\phi^6$ potential $V(\phi) = \lambda^2\phi^2(\phi^2-a^2)^2$ (dotted). (b) The kink solutions (\ref{7.2}) (top curve, blue online) and (\ref{7.4}) (bottom curve, red online) connecting $0$ to $a$, and the corresponding $\phi^6$ kink $\phi(x) = a/\sqrt{1+e^{-2\sqrt{2}a^2\lambda x}}$ (dotted). In all panels, $a=3/4$ and $b=1$.}
\label{fig:phi8_3Degen_kinks}
\end{figure}

As an illustration, consider $a=3/4$ and $b=1$. This kink, as well as the one from the previous
subsubsection, are illustrated in Fig.~\ref{fig:phi8_3Degen_kinks}. Note that, in both cases,
the kinks are asymmetric as shown by the asymptotic expressions given in
(\ref{7.2asymp}) and (\ref{7.4asymp}). For Case I, the mismatch between the $\phi^8$ and the
$\phi^6$ kink is mainly due to the slow algebraic decay (as $x\to-\infty$) of the tail of the kink, see \eqref{7.2asymp}.

\subsection{Two Degenerate Minima}

A $\phi^8$ potential with two degenerate minima can have two possible forms.
In each case, there exists a kink solution connecting the degenerate minima at $\phi=\pm a$,
as $x$ goes from $-\infty$ to $+\infty$.

\subsubsection{Case I: $\alpha_2=0$}
Let
\be\label{7.7}
V(\phi) = \lambda^2 (\phi^2-a^2)^4,
\ee
which has degenerate minima at $\phi= \pm a$. Note that in this case
$\alpha_{6,4,2,0} > 0$. The kink solution is given implicitly by
\be\label{7.8}
\mu x = \frac{2a\phi}{a^2-\phi^2}+\ln \left(\frac{a+\phi}{a-\phi}\right),
\ee
where $\mu = 4\sqrt{2} \lambda a^3$. From (\ref{7.8}), the approach to the 
asymptotes at $\phi = \pm a$ can be shown to be algebraic: 
\be
\phi(x) \simeq 
\begin{cases}
-a-\displaystyle \frac{a}{\mu x},\quad &x\to -\infty,\\[3mm]
+a-\displaystyle\frac{a}{\mu x},\quad &x\to +\infty,
\end{cases}
\ee
from which it follows that this kink is symmetric.
The corresponding kink energy is 
\be
E_k = \frac{16 \sqrt{2}}{15} \lambda a^5.
\ee

\subsubsection{Case II: $\alpha_2 > 0$}
Let
\be\label{7.5}
V(\phi) = \lambda^2 (\phi^2-a^2)^2 (\phi^2+b^2)^2,
\ee
which has degenerate minima at $\phi= \pm a$. In this case,
\be
\begin{aligned}
\alpha_{6} &= 2(b^2-a^2),\\
\alpha_{4} &= b^4 - 4a^2b^2 + a^4,\\
\alpha_{2} &= 2a^2b^2(b^2-a^2),\\
\alpha_{0} &= a^4b^4.
\end{aligned}
\ee
Clearly, $\alpha_{6,2,0} > 0$ for $b>a$, while $\alpha_{4} > 0$ as long as $b\sqrt{2-\sqrt{3}} > a$.

The kink solution is given implicitly by
\be\label{7.6}
\mu x = \frac{2a}{b} \tan^{-1} \left(\frac{\phi}{b}\right)
+\ln \left(\frac{a+\phi}{a-\phi}\right),
\ee
where $\mu = 2\sqrt{2} \lambda a (b^2+a^2)$.
From (\ref{7.6}), the approach to the 
asymptotes at $\phi = \pm a$ can be shown to be exponential:
\be
\phi(x) \simeq 
\begin{cases}
-a + 2a\, e^{\mu x + ({2a}/{b})\tan^{-1}\left(a/b\right)},\quad &x\to -\infty,\\[2mm]
+a - 2a\, e^{-\mu x + ({2a}/{b})\tan^{-1}\left(a/b\right)},\quad &x\to +\infty,
\end{cases}
\ee
from which it follows that this kink is symmetric.
The corresponding kink energy is  
\be
E_k = \frac{4\sqrt{2} }{15} \lambda a^3(a^2+5b^2).
\ee
 
\begin{figure}[h]
\centerline{\includegraphics[width=0.5\textwidth]{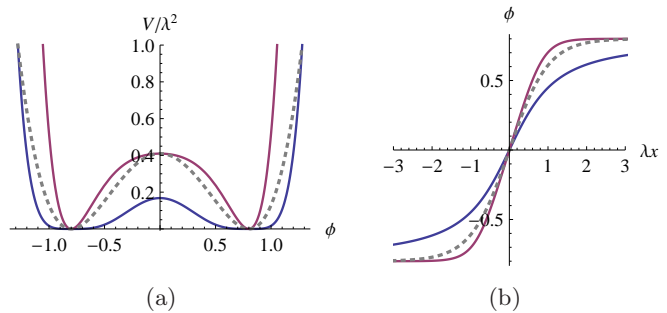}}
\hspace{1.65cm}(a)\hfill(b)\hspace{2cm}
\caption{(Color online.) $\phi^8$ field theory with two degenerate minima. (a) The potentials (\ref{7.7}) (bottom curve, blue online), (\ref{7.5}) (top curve, red online) and a representative $\phi^4$ potential $V(\phi) = \lambda^2(\phi^2-a^2)^2$ (dotted). (b) The corresponding kinks (\ref{7.8}) (inner curve, blue online) and (\ref{7.6}) (outer curve, red online)  connecting $-a$ to $+a$, and the corresponding $\phi^4$ kink $\phi(x) = a \tanh(\lambda x)$ (dotted). In all panels, $a=4/5$ and $b=1$.}
\label{fig:phi8_2Degen_kinks}
\end{figure}

%As an illustration, consider $a=b=1$ which corresponds to $\alpha_{6,2} =0$ 
%while $\alpha_{4} < 0$. 
This kink, as well as the one from the previous
subsubsection, are illustrated in Fig.~\ref{fig:phi8_2Degen_kinks}.

\subsection{Phonons}
\label{sec:phi8-phonons}

Although we have considered, without loss of generality, only
stationary kink solutions, phonon modes superimposed onto the kinks
or the equilibrium states (vacua) can be time dependent. Therefore, 
to study phonons, we must consider the nonlinear Klein--Gordon equation 
of motion for the field \cite{lohe,BishopPhysD}:
\begin{equation}
\square\phi = -V'(\phi),
\end{equation} 
where $\square \equiv \partial^2/\partial t^2 - \partial^2/\partial x^2$ is the 
d'Alembertian operator. This equation can be linearized
about any of the equilibrium states $\phi_e$ discussed above 
(e.g., $\phi_e=0$, $\phi_e=\pm a$, etc.)
to obtain an equation for the perturbation $\tilde\phi$ \cite{BishopPhysD}:
\begin{equation}
\square \tilde\phi = -V''(\phi_e)\tilde\phi.
\end{equation}
Now, seeking harmonic solutions of the form 
$\tilde\phi(x,t) \propto e^{i(qx-\omega_q t)}$, we arrive at the dispersion
relation
\be\label{eq:phi8-dispersion}
\omega_q^2 -q^2 = V''(\phi_e)
\ee
for phonon modes.
Table~\ref{table:phi8-phonos} summarizes the possible right-hand sides (RHS) in the 
dispersion relation \eqref{eq:phi8-dispersion} 
for the $\phi^8$ field theories with kink solutions studied above.
Since, $b>a$ (strictly) by assumption, cases in Table~\ref{table:phi8-phonos} for
which $V''(\phi_e)\ne0$ represent field theories with only an optical phonon branch,
while for cases with $V''(\phi_e)=0$ there is only an acoustic phonon branch. The
latter case indicates the possibility of \emph{nonlinear phonons}. This is a distinguishing 
feature of higher-than-sixth-order field theories.

\begin{table}[h]
\caption{\label{table:phi8-phonos}Phonon modes of $\phi^8$ field theory.
DM = degenerate minima. RHS = dispersion relation right-hand side.}
\begin{ruledtabular}
\begin{tabular}{lll}
potential, $V$ & equilibrium, $\phi_e$ & RHS, $V''(\phi_e)$\\
\hline
4 DM, Eq.~\eqref{4} & $\pm a$ & $8 \lambda^2 a^2 (b^2-a^2)^2$\\
4 DM, Eq.~\eqref{4} & $\pm b$ & $8 \lambda^2 b^2 (b^2-a^2)^2$\\
\hline
3 DM, Eq.~\eqref{7.1} & $\pm a$ & $8\lambda^2a^6$\\
3 DM, Eq.~\eqref{7.1} & $0$ & $0$\\
\hline
3 DM, Eq.~\eqref{7.3} & $\pm a$ & $8\lambda^2a^4(b^2+a^2)$\\
3 DM, Eq.~\eqref{7.3} & $0$ & $2\lambda^2 a^4 b^2$\\
\hline
2 DM, Eq.~\eqref{7.7} & $\pm a$ & $0$\\
\hline
2 DM, Eq.~\eqref{7.5} & $\pm a$ & $8\lambda a^2(b^2+a^2)^2$
\end{tabular}
\end{ruledtabular}
\end{table}

\section{$\phi^{10}$ Field Theory}

\subsection{The Various Phases}

The $\phi^{10}$ potential (free energy) is given, generically, by
\be\label{1.1}
V(\phi) = \lambda^2 (\phi^{10}-\alpha_8\phi^{8}+\alpha_6\phi^{6}
-\alpha_4\phi^{4}+\alpha_2\phi^2-\alpha_0),
\ee
where, 
%unless otherwise specified, $\alpha_{8,6,4,2} >0$ and, 
without loss of generality, we assume 
the coefficient of $\phi^{10}$ to be +1 in units of $\lambda^2$. 
The coefficients of $\phi^{8,6,4,2}$ are, in general, arbitrary, and
there are sixteen different possibilities, depending on whether all four, 
three, two, one or none of the coefficients are positive. However,  
if one wants to consider a model describing a succession of two 
first-order transitions then one must take $\alpha_{8,6,4,2}>0$ in 
(\ref{1.1}). As before, $\alpha_0$ in (\ref{1.1}) is 
chosen so that the minimum value of the
potential is zero, i.e., $\min_\phi V(\phi) = 0$.

While there are four parameters ($\alpha_{8,6,4,2}$) 
describing the potential, it 
can be shown, by scaling arguments, that only three of them are truly 
independent. It may be noted here that even after taking 
$\alpha_{8,6,4,2} > 0$ in (\ref{1.1}), since there are three free
parameters, there is more than one possible ``path'' to 
describing successive phase transitions. For example, one 
possible path is to start from a potential with five degenerate 
minima at $\phi=0,\pm a, \pm b$,  which is given by
\be\label{1.2}
V(\phi)= \lambda^2 \phi^2\,(\phi^2-a^2)^2(\phi^2-b^2)^2. 
\ee
As in the $\phi^{8}$ case, without any loss of generality, 
we choose $b > a$ throughout this section unless specified otherwise.
Now, what happens as $\alpha_2$ (i.e., coefficient of $\phi^2$) 
is slowly increased or decreased from this
critical value (at five degenerate minima)?  
One finds that when $\alpha_2$ is increased from this critical 
value, then  
$\phi=0$ is always the absolute minimum while the minima at 
$\phi=\pm a,\pm b$ are only local minima. On the other hand, 
if $\alpha_2$
is decreased from this critical value, then one finds that the potential has
absolute minima at $\phi=\pm b$, while the minima at 
$\phi=\pm a$ and at $\phi=0$ are now local minima. 
Thus, even with $\alpha_{8,6,4,2}>0$ in (\ref{1.1}), if one starts
with five degenerate minima, then one does not get two 
first-order transitions in succession. 

\begin{figure*}%
\centering
\subfloat[]{\includegraphics[width=0.475\textwidth]{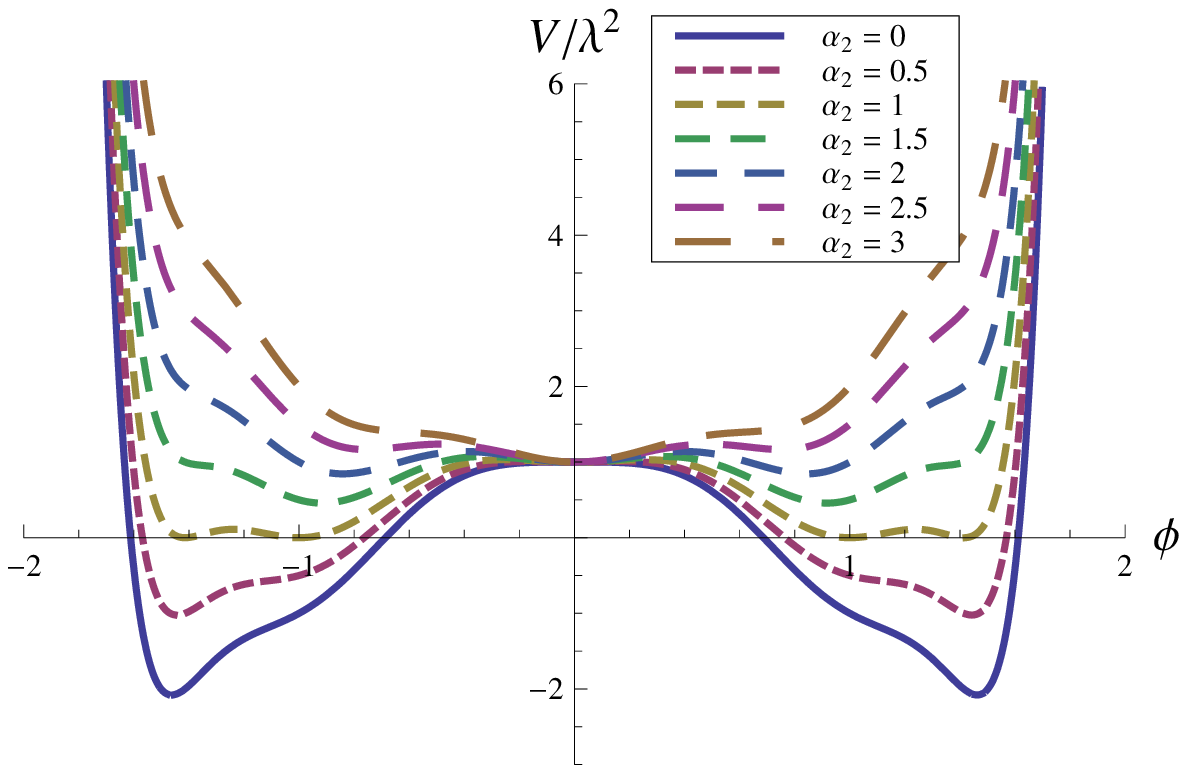}}\hfill
\subfloat[]{\includegraphics[width=0.475\textwidth]{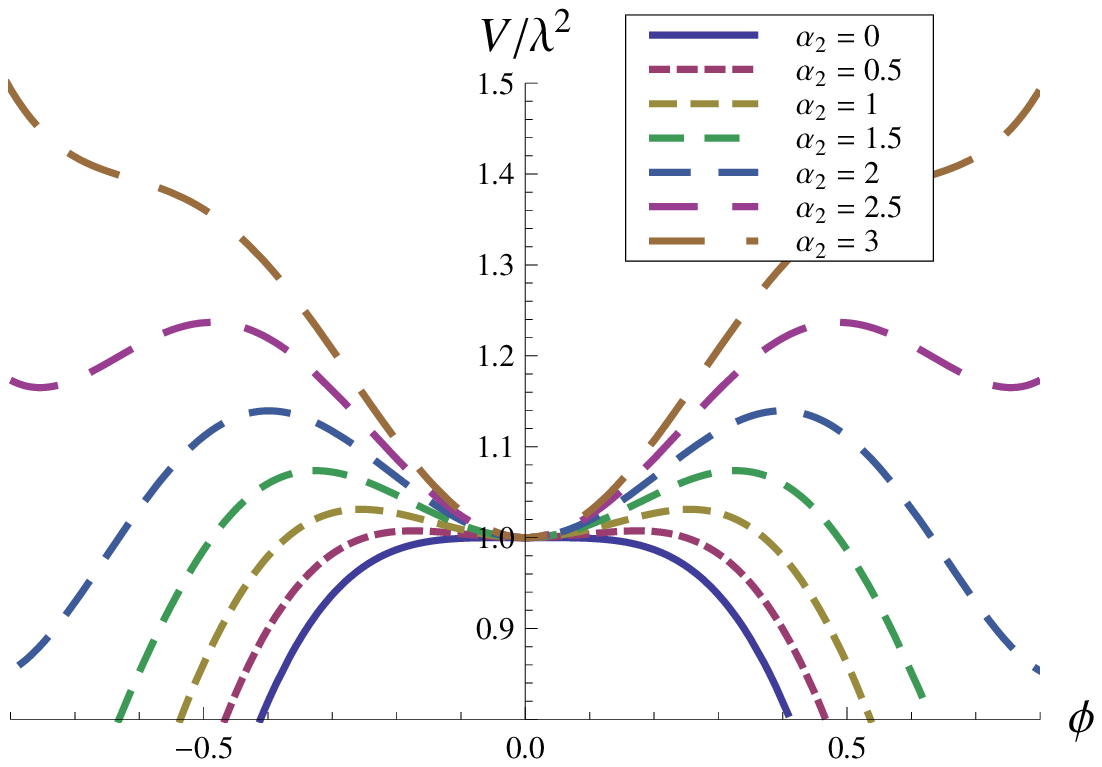}}
\caption{(Color online.) (a) Example potentials of the form (\ref{1.4}) for various illustrative values of the coefficient of the quadratic term, $\alpha_2$, showing the various phases and phase transitions in the $\phi^{10}$ theory. (b) Zoom-in of (a) near the origin.}
\label{fig:phi10_fig1}
\end{figure*}

However, if instead we start with a potential with $\alpha_{8,6,4,2}>0$ 
but with four degenerate minima,  given by 
\be\label{1.3}
V(\phi)= \lambda^2 (\phi^2+c^2)(\phi^2-a^2)^2(\phi^2-b^2)^2,
\ee
and now we vary $\alpha_2$, there are indeed two successive first-order
transitions. For potentials of this form, \eqref{1.3}, 
there are three parameters, i.e., $a$, $b$, $c$. The four coefficients 
$\alpha_{8,6,4,2}$ of the potential can be expressed in terms of the parameters
$a$, $b$, $c$ as
\be\label{eq:alpha_phi10_4degen}
\begin{aligned}
\alpha_{8} &= 2(b^2 + a^2) - c^2,\\
\alpha_{6} &= a^4 + b^4 + 4 a^2 b^2 - 2 c^2 (b^2 + a^2),\\
\alpha_{4} &= 2 a^2 b^2(b^2 + a^2) - c^2(a^4 + 4 a^2 b^2 + b^4),\\
\alpha_{2} &= a^4 b^4 - 2 a^2 b^2 c^2(b^2 + a^2),\\
\alpha_{0} &= -a^4 b^4 c^2.
\end{aligned}
\ee

Since the generic $\phi^{10}$ potential \eqref{1.1} is described by the 
four coefficients $\alpha_{8,6,4,2}$, there must exist extra constraints 
on the coefficients $\alpha_{8,6,4,2}$ to ensure a unique mapping from 
$a,b,c$ to $\alpha_{8,6,4,2}$ [recall the constraints \eqref{2} and
\eqref{4a} derived in Section~\ref{sec:phi8_var_phas}]. To this end, let 
\be
\tilde{a} = b^2+a^2,\quad \tilde{b}^2 = a^2b^2,\quad \tilde{c} = c^2,
\ee
then \eqref{eq:alpha_phi10_4degen} becomes
\be
\begin{aligned}\label{eq:alpha_phi10_4degen2}
\alpha_{8} &= 2\tilde{a} - \tilde{c},\\
\alpha_{6} &= \tilde{a}^2 +2\tilde{b}^2-2\tilde{a}\tilde{c},\\
\alpha_{4} &= 2\tilde{a}\tilde{b}^2-\tilde{c}(\tilde{a}^2+2\tilde{b}^2),\\
\alpha_{2} &= \tilde{b}^2(\tilde{b}^2-2\tilde{a}\tilde{c}),\\
\alpha_{0} &= -\tilde{b}^2 \tilde{c}.
\end{aligned}
\ee
By definition, $\tilde{a}^2 > 4\tilde{b}^2$ [i.e., $(a^2+b^2)^2 > 4a^2 b^2$ or, equivalently $(b^2-a^2)^2>0$], hence
\be
\begin{aligned}
4\tilde{b}-\tilde{c} &< \alpha_8 < 2\tilde{a},\\
2(3\tilde{b}^2-\tilde{a}\tilde{c}) &< \alpha_6 < \tilde{a}^2+2\tilde{b}^2 -4\tilde{b}\tilde{c},\\
4\tilde{b}^3 -\tilde{a}^2 \tilde{c} &< \alpha_4 < 2(\tilde{a}-3\tilde{c})\tilde{b}^2,\\
\tilde{b}^4 - \tilde{c}\tilde{a}^3/2 &< \alpha_2 < (\tilde{a}\tilde{b}^2/4)(\tilde{a}-8\tilde{c}),
\end{aligned}
\ee
where $\tilde{b}$ is the positive root of $\tilde{b}^2 = a^2b^2$. This
set of inequalities provides the signs of $\alpha_{8,6,4,2}$ in terms of $a,b,c$.
Furthermore, we note that $\tilde{b}$ and $\tilde{c}$ can be eliminated
between the first four equations in \eqref{eq:alpha_phi10_4degen2} to
obtain
\be
\begin{aligned}
4 \alpha_2 + (5 \tilde{a}^2 - \alpha_6 - 2 \tilde{a} \alpha_8) (3 \tilde{a}^2 + \alpha_6 - 2 \tilde{a} \alpha_8) &= 0,\\
5 \tilde{a}^3 + \alpha_4 + \tilde{a} \alpha_6 + 2 \tilde{a} \alpha_8^2 - (6 \tilde{a}^2 + \alpha_6) \alpha_8 &= 0.
\end{aligned}
\ee
Then, it is possible to eliminate $\tilde{a}$ between the last two equations
to obtain the desired constraint [analogue of \eqref{2} for the $\phi^8$ field
theory with four degenerate minima]:
\begin{multline}
8000 \alpha_2^3 + (27 \alpha_4^2 + 4 \alpha_6^3 - 
     18 \alpha_4 \alpha_6 \alpha_8 - \alpha_6^2 \alpha_8^2 + 
     4 \alpha_4 \alpha_8^3)\\
\times (25 \alpha_4^2 - 20 \alpha_6^3 - 
     70 \alpha_4 \alpha_6 \alpha_8 + 37 \alpha_6^2 \alpha_8^2 + 
     4 \alpha_4 \alpha_8^3 - 8 \alpha_6 \alpha_8^4)\\
+ 8 \alpha_2 \Big[15 \alpha_4^2 (15 \alpha_6 + 26 \alpha_8^2)
+ 2 \alpha_4 (125 \alpha_6^2 \alpha_8 - 262 \alpha_6 \alpha_8^3 + 56 \alpha_8^5)\\
+ (4 \alpha_6 - \alpha_8^2) (35 \alpha_6^3 - 66 \alpha_6^2 \alpha_8^2 + 48 \alpha_6 \alpha_8^4 - 
        8 \alpha_8^6) \Big]\\
= 16 \alpha_2^2 (325 \alpha_6^2 + 600 \alpha_4 \alpha_8 - 440 \alpha_6 \alpha_8^2 + 88 \alpha_8^4).
\end{multline}
This constraint ensures that $a,b,c$ can be uniquely mapped to $\alpha_{8,6,4,2}$.

As an illustration, in Fig.~\ref{fig:phi10_fig1}, we have plotted the potential 
\be\label{1.4}
V(\phi)=\lambda^2 [\phi^{10} -5.75\phi^8+ 11.5\phi^6 -8.75 \phi^4 
+ \alpha_2 \phi^2 +1],
\ee
for various values of the parameter $\alpha_2$, in units of $\lambda^2$,
to illustrate the structure of the phases. For
$\alpha_2=1=\alpha_2^{c}(II)$ this has four 
degenerate minima [this is the second first-order transition point, 
i.e., $T=T_c^{I}(II)$]. 
In particular, when $\alpha_2=1$, the potential (\ref{1.4})
is of the form (\ref{1.3}) with $a=1$, $b=\sqrt{2}$, $c=1/4$.

If the temperature is  increased slightly above $T^{I}_c (II)$, 
i.e., $\alpha_2$ is increased slightly beyond $\alpha_2^{c}=1$, then the potential \eqref{1.4}
has two absolute minima at $\phi = \pm \hat{a}$ $(\hat{a}<a= 1)$, local
minima at $\phi=0, \pm \hat{b}$ $(\hat{b}<b=\sqrt{2}$),
and there are four maxima between
them. As $\alpha_2$ is further increased [i.e., $T$ is
further increased beyond $T^{I}_c (II)$], 
there comes a point [$\alpha_2^{c}(I)=2.2$ for the potential \eqref{1.4}], 
at which the potential has degenerate minima at 
$\phi=0$ and at $\phi=\pm a$. Thus,
this is the first first-order transition point $T_c^{I} (I)$. 
This is because, if the temperature is 
increased beyond this critical value [i.e., if $\alpha_2$ is further 
increased beyond $\alpha_2^{c}(I)$],
then $\phi=0$ becomes the absolute minimum,
while the minima at $\phi=\pm a$ disappear completely.
 
As far as the two local minima at $\phi = \pm b$ are concerned,
they disappear at some point as the temperature is increased 
beyond $T_c^{I} (II)$, with the precise value of $\alpha_2$ 
depending on the values of the other parameters 
[in Fig.~\ref{fig:phi10_fig1} they disappear 
at $\alpha_2=4$, much above $T_c^{I} (I)$]. 

If, instead, $\alpha_2$ is decreased from $\alpha_2^{c}(II)=1$,
[i.e., temperature is lowered below $T^{I}_c (II)$], 
then the potential has two absolute minima
at $\phi = \pm \hat{b}$ $(\hat{b}>b=\sqrt{2})$, local minima at $\phi=0,\pm
\hat{a}$ $(\hat{a}>a=1$), and there are four maxima between them.
As the temperature is further lowered so that $\alpha_2$ approaches zero, 
the local minima at $\phi = \pm \hat{a}$ 
disappear. For $\alpha_2 \le 0$, the potential only has 
two minima at $\phi=\pm \hat{b}$ $(\hat{b} < b = \sqrt{2})$, a maximum at $\phi=0$,
and this picture persists, no matter how much further the temperature is lowered.  

It is insightful to note that the structure near the
first first-order transition point $T_c^{I} (I)$ is similar to that in the 
$\phi^6$ model for a first-order phase transition \cite{bk}. 
Meanwhile, the structure near the second first-order transition point
$T_c^{I}(II)$ is similar to that of the asymmetric double well $\phi^4$ 
model of a first-order  phase transition \cite{sanati}.

\subsection{Five Degenerate Minima}

Consider the $\phi^{10}$ potential given in (\ref{1.2}). In this case,
\be\label{1.12a}
\begin{aligned}
\alpha_{8} &= 2(b^2+a^2),\\
\alpha_{6} &= a^4+b^4+4a^2b^2,\\
\alpha_{4} &= 2a^2 b^2 (b^2+a^2),\\
\alpha_{2} &= a^4 b^4,\\
\alpha_{0} &= 0.
\end{aligned}
\ee
Clearly, $\alpha_{8,6,4,2}$ are strictly positive.
This potential has five degenerate minima at $\phi=0,\pm a,\pm b$,
and, hence, four kink solutions exist, only two of which are distinct
due to the symmetry of the potential. 

\subsubsection{Kink connecting $0$ to $a$ (or $-a$ to $0$)}
This kink solution is given implicitly by
\be\label{1.15}
e^{\mu x} = \frac{\phi^{2(\gamma-1)} (b^2-\phi^2)}
{(a^2-\phi^2)^{\gamma}},
\ee
where $\mu = 2\sqrt{2}\lambda b^2(b^2-a^2)$ and $\gamma = b^2/a^2$ 
($>1$ by assumption). From \eqref{1.15}, the approach to the asymptotes
at $\phi = 0,a$ can be shown to be exponential:
\be
\phi(x) \simeq \begin{cases}
\displaystyle\frac{a^{\gamma/(\gamma-1)}}{b^{1/(\gamma-1)}}e^{\mu x/[2(\gamma-1)]}, &x\to -\infty,\\[3mm]
a - \displaystyle\frac{a(b^2-a^2)^{1/\gamma}}{2a^{2/\gamma}}e^{-\mu x/\gamma}, &x\to +\infty.
\end{cases}
\ee
%Analogous expressions can be derived for the asymptotes at $\phi=-a,0$.
Note, however, that the rate at which $\phi$ asymptotes to $0$ is given by $\mu/[2(\gamma-1)]$,
while the rate at which $\phi$ asymptotes to $a$ is given by $\mu/\gamma$,
hence this kink is {asymmetric}.
The corresponding kink energy is 
\be\label{1.15e}
E_k^{(1)} = \frac{\sqrt{2}}{12} \lambda a^4(3b^2-a^2).
\ee

\begin{figure*}
\centerline{\includegraphics[width=0.8\textwidth]{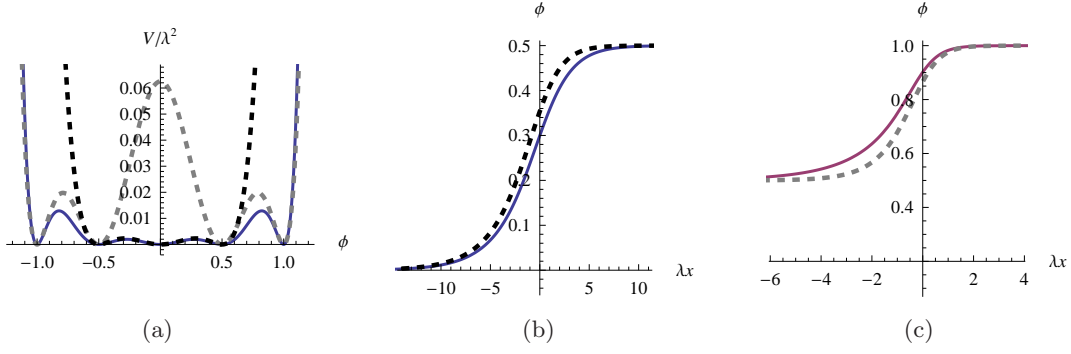}}
\hspace{3.6cm}(a)\hfill(b)\hfill(c)\hspace{3.8cm}
\caption{(Color online.) $\phi^{10}$ field theory with five degenerate minima. (a) The potential (\ref{1.2}) (solid), a representative $\phi^8$ potential with four degenerate minima \eqref{4} (gray, dotted) and a representative $\phi^6$ potential with three degenerate minima $V(\phi) = \lambda^2\phi^2(\phi^2-a^2)^2$ (black, dotted). (b) The kink solution (\ref{1.15}) (solid) connecting $0$ to $a$ and the corresponding $\phi^6$ kink $\phi(x) = a/\sqrt{1+e^{-2\sqrt{2}a^2\lambda x}}$ (black, dotted). (c) The kink solution (\ref{1.15d}) (solid) connecting $a$ to $b$ and the corresponding $\phi^8$ kink  \eqref{4.2b} (gray, dotted). In all panels, $a=1/2$ and $b=1$.}
\label{fig:phi10_5Degen_kinks}
\end{figure*}

\subsubsection{Kink connecting $a$ to $b$ (or $-b$ to $-a$)}
In this case, the kink solution is given implicitly by
\be\label{1.15d}
e^{\mu x} = \frac{(\phi^2-a^2)^{\gamma}}{\phi^{2(\gamma-1)} (b^2-\phi^2)},
\ee
where $\mu = 2\sqrt{2}\lambda b^2(b^2-a^2)$ and $\gamma = b^2/a^2$.
From \eqref{1.15d}, the approach to the asymptotes
at $\phi = a,b$ can be shown to be exponential: 
\be
\phi(x) \simeq \begin{cases}
a+\displaystyle\frac{(b^2-a^2)^{1/\gamma}}{2a^{(2-\gamma)/\gamma}} e^{\mu x/\gamma}, &x\to -\infty,\\[3mm]
b-\displaystyle\frac{(b^2-a^2)^\gamma}{2b^{2\gamma-1}}e^{-\mu x}, &x\to +\infty.
\end{cases}
\ee
%Analogous expressions can be derived for the asymptotes at $\phi=-b,-a$.
Note, however, that the rate at which $\phi$ asymptotes to $a$ is given by $\mu/\gamma$,
while the rate at which $\phi$ asymptotes to $b$ is given by $\mu$,
hence this kink is also {asymmetric}.
The corresponding kink energy is

\be\label{1.15f}
E_k^{(2)} = \frac{\sqrt{2}}{12} \lambda (b^2-a^2)^3.
\ee

Note that $E_k^{(1)} \gtreqqless E_k^{(2)}$ for $b/a \lesseqqgtr \sqrt{3}$.
As for the similar $\phi^8$ case (Section~\ref{sec:phi8-tcI-a}), 
it would be of interest to 
study the interaction energy between two kinks of the same type as well 
as two kinks of different types but with equal energies.

As an illustration, consider $a=1/2$ and $b=1$. This kink, as well as the one from the previous
subsubsection, are illustrated in Fig.~\ref{fig:phi10_5Degen_kinks}.  Since the potential (\ref{1.2})
has five degenerate minima, it is possible to fit a $\phi^8$ potential with four degenerate
minima (at $\phi = \pm a$ and $\phi = \pm b$) to it, and also a $\phi^6$ potential with
three degenerate minima (at $\phi = 0$ and $\phi = \pm a$). As can be seen in
Fig.~\ref{fig:phi10_5Degen_kinks}, for the parameters chosen,
the shapes of the corresponding kink solutions from the lower-order field 
theories closely match those of the $\phi^{10}$ theory.

\begin{widetext}
\subsection{Four Degenerate Minima}

\subsubsection{$T=T_c^I(II)$}

Consider the $\phi^{10}$ potential given in \eqref{1.3}.
This potential has four degenerate minima at $\phi = \pm a, \pm b$,
and, hence, three kink solutions exist, only two of which are distinct
due to the symmetry of the potential.

\paragraph{Kink connecting $-a$ to $+a$}
This kink solution is given implicitly by
\be\label{1.5b}
\mu x =  \bigg \{\sinh^{-1}
\left[\frac{c+\alpha \phi}{\alpha(a-\phi)}\right] - \sinh^{-1}
\left[\frac{c-\alpha \phi}{\alpha(a+\phi)}\right] \bigg \}
+\frac{\alpha\sqrt{1+\alpha^2}}{\beta\sqrt{1+\beta^2}}  \bigg \{\sinh^{-1}
\left[\frac{c-\beta \phi}{\beta(b+\phi)}\right]- \sinh^{-1}
\left[\frac{c+\beta \phi}{\beta(b-\phi)}\right] \bigg \},
\ee
where $\mu =2\sqrt{2} \lambda \alpha\sqrt{1+\alpha^2} 
(\beta^2-\alpha^2)c^4$,
$\beta = b/c$ and $\alpha = a/c$
with $\beta > \alpha$ by assumption.
From \eqref{1.5b}, the approach to the asymptotes
at $\phi = \pm a$ can be shown to be exponential: 
\be
\phi(x) \simeq \begin{cases}
-a+\frac{2(c+\alpha a)}{\alpha} \exp\left( \sinh^{-1}\left[\frac{1}{2}(1 - \alpha^{-2})\right]+\frac{\alpha  \sqrt{1+\alpha ^2} }{\beta  \sqrt{1+\beta ^2}} \left\{\sinh^{-1}\left[\frac{c-\beta a}{\beta(b + a)}\right] -\sinh^{-1}\left[\frac{c+\beta a}{\beta(b-a)}\right]\right\}\right) e^{\mu x}, &x\to -\infty,\\[3mm]
+a-\frac{2(c+\alpha a)}{\alpha} \exp\left( \sinh^{-1}\left[\frac{1}{2}(1 - \alpha^{-2})\right]+\frac{\alpha  \sqrt{1+\alpha ^2} }{\beta  \sqrt{1+\beta ^2}} \left\{\sinh^{-1}\left[\frac{c-\beta a}{\beta(b + a)}\right] -\sinh^{-1}\left[\frac{c+\beta a}{\beta(b-a)}\right]\right\}\right) e^{-\mu x}, &x\to +\infty.
\end{cases}
\ee
Clearly, this kink is symmetric. 
%[as can also be easily verified by noting \eqref{1.5b} is invariant under the transformation $\{x,\phi\}\mapsto\{-x,-\phi\}$]
The kink's energy is
\be\label{1.5f}
E_k^{(1)} = \frac{\sqrt{2}}{{24}} \lambda \bigg \{ \alpha\sqrt{1+\alpha^2} 
(12a^2 b^2 - 4a^2 c^2 - 6b^2 c^2 -4a^4- 3c^4 )
+3 c^2
[8a^2 b^2+2(b^2 + a^2) c^2 +c^4]\sinh^{-1}\alpha \bigg \}.
\ee

\begin{figure*}
\centerline{\includegraphics[width=0.8\textwidth]{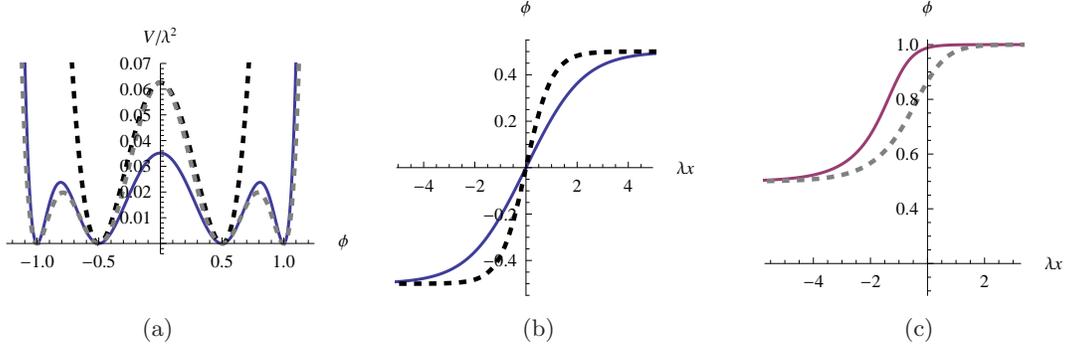}}
\hspace{3.6cm}(a)\hfill(b)\hfill(c)\hspace{3.8cm}
\caption{(Color online.) $\phi^{10}$ field theory at the second first-order phase transition, $T=T_c^I(II)$. (a) The potential (\ref{1.3}) (solid), a representative $\phi^8$ potential with four degenerate minima \eqref{4} (gray, dotted) and a representative $\phi^4$ potential $V(\phi) = \lambda^2(\phi^2-a^2)^2$ (black, dotted). (b) The kink solution (\ref{1.5b}) (solid) connecting $-a$ to $+a$ and the corresponding $\phi^4$ kink $\phi(x) = a\tanh(\lambda x)$ (black, dotted). (c) The kink solution (\ref{1.5k}) (solid) connecting $a$ to $b$ and the corresponding $\phi^8$ kink \eqref{4.2b} (gray, dotted). In all panels, $a=1/2$, $b=1$ and $c=3/4$.}
\label{fig:phi10_TCI _II _kinks}
\end{figure*}

\paragraph{Kink connecting $a$ to $b$ (or $-b$ to $-a$)}
In this case, the kink solution is given implicitly by
\be\label{1.5k}
\mu x =  -\bigg \{ \sinh^{-1}
\left[\frac{c+\alpha \phi}{\alpha(\phi-a)}\right] - \sinh^{-1}
\left[\frac{c-\alpha \phi}{\alpha(a+\phi)}\right] \bigg \}
-\frac{\alpha\sqrt{1+\alpha^2}}{\beta\sqrt{1+\beta^2}}  \bigg\{ \sinh^{-1}
\left[\frac{c-\beta \phi}{\beta(b+\phi)}\right]- \sinh^{-1}
\left[\frac{c+\beta \phi}{\beta(b-\phi)}\right] \bigg \},
\ee
where $\mu$, $\beta$ and $\alpha$ are defined below \eqref{1.5b}.
From \eqref{1.5k}, the approach to the asymptotes
at $\phi = a,b$ can be shown to be exponential: 
\be
\phi(x) \simeq \begin{cases}
a+\frac{2(c+\alpha a)}{\alpha} \exp\left( \sinh^{-1}\left[\frac{\alpha^2-1}{2\alpha^2}\right]
+ \frac{\alpha  \sqrt{1+\alpha ^2} }{\beta  \sqrt{1+\beta ^2}} \left\{\sinh^{-1}\left[\frac{c-\beta a}{\beta(b + a)}\right] -\sinh^{-1}\left[\frac{c+\beta a}{\beta(b-a)}\right]\right\}\right) e^{\mu x}, &x\to -\infty,\\[3mm]
b-\frac{2(c+\beta b)}{\beta} \exp\left( \sinh^{-1}\left[\frac{\beta^2-1}{2\beta^2}\right] + \frac{\beta  \sqrt{1+\beta ^2}}{\alpha  \sqrt{1+\alpha ^2} } \left\{\sinh^{-1}\left[\frac{c-\alpha b}{\alpha(b + a)}\right] + \sinh^{-1}\left[\frac{c+\alpha b}{\alpha(b-a)}\right]\right\}\right) e^{-\mu \frac{\beta  \sqrt{1+\beta ^2}}{\alpha  \sqrt{1+\alpha ^2} } x}, &x\to +\infty.
\end{cases}
\ee
%Analogous expressions can be derived for the asymptotes at $\phi=-b,-a$.
Note, however, that the rate at which $\phi$ asymptotes to $a$ is given by $\mu$,
while the rate at which $\phi$ asymptotes to $b$ is given by 
$\mu {\beta  \sqrt{1+\beta ^2}}/({\alpha  \sqrt{1+\alpha ^2} })$,
hence this kink is {asymmetric}.
The kink's energy is
\begin{multline}\label{1.5g}
E_k^{(2)} = \frac{\sqrt{2}}{48} \lambda \bigg \{ 
\alpha\sqrt{1+\alpha^2}
(12a^2 b^2 - 4a^2 c^2 - 6 b^2 c^2 - 4 a^4 - 3 c^4)
-\beta\sqrt{1+\beta^2}
(12a^2 b^2 - 4b^2 c^2 - 6a^2 c^2 - 4 b^4 - 3c^4)\\
+3c^2
[8a^2 b^2 + 2(b^2 + a^2) c^2 +c^4] \left(\sinh^{-1}\alpha
- \sinh^{-1}\beta\right) \bigg \}.
\end{multline}

\end{widetext}

Figure~\ref{fig:phi10_TCI _II _kinks} shows the kink solutions from the previous
subsubsection. Note that, unlike the $\phi^8$ case in Fig.~\ref{fig:phi8_TCI_kinks},
the match between the $\phi^{10}$ and $\phi^4$ theories for the symmetric kink
connecting $-a$ to $+a$ is not very good for the chosen parameters. 
The agreement between the two is determined by the curvature of the potential 
near $\phi=0$, which is controlled by $c$; 
for other values of $c$, these can be made more similar. Specifically, as $c\to1$
the $\phi^{10}$ and $\phi^4$ kinks match well [Fig.~\ref{fig:phi10_TCI _II _kinks}(b)], 
while as $c\to0$, the $\phi^{10}$ and $\phi^8$ kinks match better 
[Fig.~\ref{fig:phi10_TCI _II _kinks}(c].

\subsubsection{$T_c^{I}(II)<T<T_c^{I} (I)$}

For temperatures between the two first-order phase transitions
[i.e., $1 < \alpha_2 < 2.2$ for the example potential
(\ref{1.4})], the potential can be rewritten as
\be\label{1.6}
V(\phi) = \lambda^2 (\phi^2 -\hat{a}^2)^2 (\phi^2+c^2)
[\phi^4 - d \phi^2 + e],
\ee  
with $\hat{a}^2 <a^2$ and $d^2 < 4e$ so that the 
minimum of the potential is indeed at 0.
We expect a kink solution exists connecting the two degenerate minima
$\phi =\pm\hat{a}$, as $x$ goes from $-\infty$ to $+\infty$.

As an illustration, consider the factorized potential
\be\label{1.7}
V(\phi) = \lambda^2 (\phi^2-0.9 )^2 (\phi^2+0.2)[\phi^4-4.15 \phi^2 +4.45].
\ee
This potential has absolute minima at 
$\phi= \pm\hat{a} = \pm \sqrt{0.9}$ and local minima at $\phi=0$ and  
$\phi=\pm\hat{b}$  $(\hat{b}^2<b^2=2)$.

\subsubsection{$T<T_c^{I}(II)$}

Below the second first-order phase transition
[i.e., $ \alpha_2 < 1$ for the example potential
(\ref{1.4})], the potential can be rewritten as
\be\label{1.10}
V(\phi) = \lambda^2 (\phi^2 -\hat{b}^2)^2 (\phi^2+c^2)
[\phi^4 - d \phi^2 + e],
\ee  
with $\hat{b}^2 > b^2$ and $d^2 < 4e$ 
so that the minimum of the potential is indeed at 0. 
We expect a kink solution exists connecting the degenerate minima
$\phi =\pm\hat{b}$, as $x$ goes from $-\infty$ to $+\infty$.

As an illustration, consider the factorized potential
\be\label{1.11}
V(\phi) = \lambda^2 (\phi^2-2.05 )^2 (\phi^2+0.3) [\phi^4-1.97 \phi^2+1.15].
\ee
This potential has absolute minima at 
$\phi = \pm\hat{b} = \pm\sqrt{2.05}$ and local minima at $\phi=0$ and at 
$\phi=\pm\hat{a}$ $(\hat{a}^2>a^2=1)$.

\begin{figure}[h]
\centerline{\includegraphics[width=0.5\textwidth]{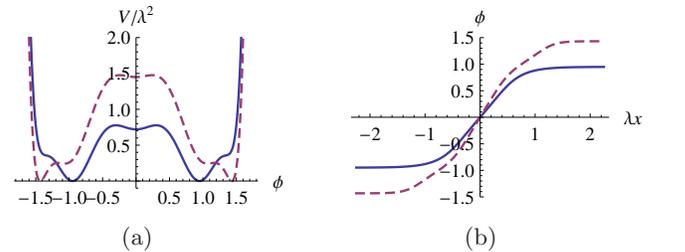}}
\hspace{1.65cm}(a)\hfill(b)\hspace{2cm}
\caption{(Color online.) Kink solutions between the first and second first-order phase transition [$T_c^I(II)<T<T_c^I(I)$] and below the second first-order phase transition [$T < T_c^I(II)$] in $\phi^{10}$ field theory. (a) The potentials (\ref{1.7}) (solid) and (\ref{1.11}) (dashed). (b) The corresponding kinks computed by solving the equation of motion $d\phi/dx = \sqrt{2V(\phi)}$ numerically subject to the symmetry condition $\phi(0) = 0$.}
\label{fig:phi10_nonTCI_II_kinks}
\end{figure}

This kink solution for $T<T_c^{I}(II)$  and the previous one 
for $T_c^{I}(II)<T<T_c^{I} (I)$ are illustrated in Fig.~\ref{fig:phi10_nonTCI_II_kinks}. 
Notice that for the case $T<T_c^I(II)$ [dashed curve in 
Fig.~\ref{fig:phi10_nonTCI_II_kinks}(b)], the kink ``feels'' 
the influence of the two local minima at $\phi \approx \pm 1.17101$, 
similarly to kinks in certain cases of $\phi^6$ field theory \cite{sanati99}, and the kink near 
the first-order phase transition in $\phi^8$ field 
theory (recall Section~\ref{sec:phi8-t_less_TcI}).
However, for these choices of $d$ and $e$,
neither set of kinks  in Fig.~\ref{fig:phi10_nonTCI_II_kinks}
appears to ``feel'' the influence of the local minimum at $\phi = 0$.

\subsection{Three Degenerate Minima}

There are four possible forms of the $\phi^{10}$ potential with three degenerate minima
for which kink solutions can be constructed.
These potentials have three degenerate minima,
and, hence, two kink solutions exist, only one of which is distinct
due to the symmetry of the potential.

\subsubsection{Case I}
%with degenerate minima at $\phi=0$
%and at $\phi^2= \hat{a}^2<a^2$. In this case,
%one can always rewrite (\ref{1.1}) as
%\be\label{1.8}
%V(\phi) = \lambda^2 \phi^2 (\phi^2 -\hat{a}^2)^2 
%[\phi^4 - \gamma \phi^2 +\delta],
%\ee  
%with $\gamma > 0$, $\hat{a}^2 <a^2=1$ and 
%$4\delta > \gamma^2$ so that the 
%minima of the potential are at $\phi=0$ and $\pm \hat{a}$. 
%In this case, $\alpha_{8,6,4,2} >0$. 
%For uniformity of notation, rewrite (\ref{1.8}) as 

First, consider the potential
\be\label{1.8g}
V(\phi) = \lambda^2 \phi^2 (\phi^2 -a^2)^2 [\phi^4 - b \phi^2 + c], \quad b^2<4c,
\ee
with $b > 0$, so that the potential has degenerate minima at $\phi=0,\pm a$.
In this case,
\be
\begin{aligned}
\alpha_8 &= 2a^2 + b,\\
\alpha_6 &= a^2(a^2 + 2b) +c,\\
\alpha_4 &= a^2(a^2b+2c),\\
\alpha_2 &= a^4c,\\
\alpha_0 &= 0.
\end{aligned}
\ee
Clearly, $\alpha_{8,6,4,2}$ are strictly positive.

The corresponding kink solution connecting $0$ to $+a$ (or $-a$ to $0$) is given implicitly by
\begin{multline}\label{1.8b}
\mu x =  \frac{\sqrt{c}}{\sqrt{c+a^4}}\sinh^{-1}
\left [\frac{2c+ba^2+(2a^2-b) \phi^2}{(a^2-\phi^2) \sqrt{4a^2b-b^2+4c}} \right]\\
- \sinh^{-1} \left(\frac{2c -b \phi^2}{\phi^2 \sqrt{4c-b^2}} \right) ,
\end{multline}
where $\mu =2\sqrt{2} \lambda a^2 \sqrt{c}$. 
From \eqref{1.8b}, the approach to the asymptotes
at $\phi = 0,a$ can be shown to be exponential:
\begin{widetext}
\be
\phi(x) \simeq \begin{cases}
\frac{2\sqrt{c}}{(4c-b^2)^{1/4}} \exp\left( -\frac{\sqrt{c}}{2\sqrt{c+a^4}}\sinh^{-1}
\left [\frac{2c+ba^2}{a^2\sqrt{4a^2b-b^2+4c}} \right]  \right) e^{\mu x/2}, &x\to -\infty,\\[3mm]
a-\frac{2(c+a^4)}{a\sqrt{4a^2b - b^2 +4c}} \exp\left(-\frac{\sqrt{c+a^4}}{\sqrt{c}} \sinh^{-1}\left[\frac{2c-ba^2}{a^2\sqrt{4c-b^2}} \right] \right) e^{-\mu x {\sqrt{c+a^4}}/{\sqrt{c}}}, &x\to +\infty.
\end{cases}
\ee
\end{widetext}
%Analogous expressions can be derived for the asymptotes at $\phi=-a,0$.
Note, however, that the rate at which $\phi$ asymptotes to $0$ is given by $\mu/2$,
while the rate at which $\phi$ asymptotes to $a$ is given by 
$\mu\sqrt{c+a^4}/\sqrt{c}$, hence this kink is {asymmetric}.
The kink's energy is
\begin{multline}\label{18t}
E_k = \frac{\sqrt{2}}{96} \lambda\bigg \{ 2(3 b^2 + 4 a^4 - 4a^2b - 8c)\sqrt{a^4-ba^2+c}\\
+16c^{3/2}
%-8(a^4-ba^2+c)^{3/2} 
+6b(2a^2-b)\sqrt{c}
+3(b^2-4c)(2a^2-b)\\ \times
\ln\left[ \frac{-b+2\sqrt{c}}{2a^2-b+2\sqrt{a^4-ba^2+c}}\right] \bigg \}.
\end{multline}
%As an illustration, consider the potential
%\be\label{1.9}
%V(\phi) = \lambda^2 \phi^2 (\phi^2-0.65 )^2 [\phi^4-4.45 \phi^2+5.36],
%\ee
%which has absolute degenerate minima at $\phi=0$ and at $\phi = \pm\sqrt{0.65}$.

\subsubsection{Case II}
Now, let
\be\label{1.8c}
V(\phi) = \lambda^2 \phi^2 (\phi^2 -a^2)^2 (\phi^2+b^2)^2.
\ee  
This potential has three degenerate minima at $\phi = 0,\pm a$. 
In this case,
\be\label{1.8x}
\begin{aligned}
\alpha_{8} &= 2(b^2-a^2),\\
\alpha_{6} &= b^4+a^4-4a^2 b^2,\\
\alpha_{4} &= 2a^2 b^2 (b^2 -a^2),\\
\alpha_{2} &= a^4 b^4,\\
\alpha_{0} &= 0.
\end{aligned}
\ee
Clearly, $\alpha_{8,4,2} > 0$ for $b>a$, while $\alpha_6 > 0$ as long as $b \sqrt{2-\sqrt{3}} > a$.

The corresponding kink solution connecting $0$ to $+a$ (or $-a$ to $0$) is given implicitly by
\be\label{1.8e}
e^{\mu x} = \frac{\phi^2}{(a^2-\phi^2)^{b^2/(b^2+a^2)}
(b^2+\phi^2)^{a^2/(b^2+a^2)}},
\ee 
where $\mu =2\sqrt{2} a^2 b^2 \lambda$.
From \eqref{1.8e},
it can be shown that the approach to the asymptotes at 
$\phi = 0,a$ is exponential:
\be
\phi(x) \simeq \begin{cases}
a^{b^2/(b^2+a^2)}b^{a^2/(b^2+a^2)}e^{\mu x/2}, &x\to -\infty,\\[2mm]
a-\displaystyle\frac{a^{1+2a^2/b^2}}{2(a^2+b^2)^{a^2/b^2}}e^{-\mu (b^2+a^2)x/b^2}, &x\to +\infty.
\end{cases}
\ee
Consequently, this kink is asymmetric due to its different growth rates  
as $x \to\pm\infty$.
%Analogous expressions can be derived for the asymptotes at $\phi=-a,0$.
The kink's energy is
\be
E_k = \frac{\sqrt{2}}{12} \lambda a^4(a^2+3b^2).
\ee

\subsubsection{Case III}
Next, consider
\be\label{1.15a}
V(\phi) = \lambda^2 \phi^{6}(\phi^2-a^2)^2,
\ee
In this case $\alpha_{8,6}>0$, while $\alpha_{4,2,0}=0$.
%\be
%\begin{aligned}
%\alpha_{8} &= 2a^2,\\
%\alpha_{6} &= a^4,\\
%\alpha_{4} &= \alpha_{2} = \alpha_{0} = 0.
%\end{aligned}
%\ee
%Clearly, $\alpha_{8,6}>0$. 

The corresponding kink solution connecting $0$ to $+a$ (or $-a$ to $0$) is given implicitly by
\be\label{1.15b}
\mu x = -\frac{a^2}{\phi^2} 
+ \ln\left(\frac{\phi^2}{a^2-\phi^2}\right),
\ee
where $\mu = 2\sqrt{2}\lambda a^4$. From \eqref{1.15b},
it can be shown that the approach to the asymptotes at 
$\phi = 0,a$ is of mixed type:
\be\label{1.15b-a}
\phi(x) \simeq \begin{cases}
\displaystyle\frac{a}{\sqrt{-\mu x}}, &x\to -\infty,\\[3mm]
a-\frac{1}{2} ae^{-\mu x-1}, &x\to +\infty.
\end{cases}
\ee
Consequently, this kink is asymmetric due to the algebraic versus
exponential approach as $x \to\pm\infty$, respectively.
%Analogous expressions can be derived for the asymptotes at $\phi=-a,0$.
The kink's energy is
\be
E_k = \frac{\sqrt{2}}{12} \lambda a^6.
\ee

\subsubsection{Case IV}
Finally, consider the potential
\be\label{1.80}
V(\phi) = \lambda^2 \phi^4(\phi^2-a^2)^2 (\phi^2+b^2),
\ee
for which 
\be
\begin{aligned}
\alpha_{8} &= 2a^2-b^2,\\
\alpha_{6} &= a^2(a^2 - 2b^2),\\
\alpha_{4} &= -a^4b^2,\\
\alpha_{2} &= \alpha_{0} = 0.
\end{aligned}
\ee
In this case, $\alpha_8, \alpha_6 >0$  as long as $a > \sqrt{2} b$, and $\alpha_4 < 0$.

The corresponding kink solution connecting $0$ to $+a$ (or $-a$ to $0$) is given implicitly by
\begin{multline}\label{1.81}
\mu x = 
-\frac{2a\sqrt{b^2+a^2} \sqrt{\phi^2+b^2}}{b^2 \phi}\\
+\sinh^{-1} \left [\frac{b^2+a\phi}{b(a-\phi)} \right ]
-\sinh^{-1} \left [\frac{b^2-a\phi}{b(a+\phi)} \right ],
\end{multline}
where $\mu = 2\sqrt{2}\lambda a^3\sqrt{a^2+ b^2}$.
From \eqref{1.81},
it can be shown that the approach to the asymptotes at 
$\phi = 0,a$ is of mixed type:
\be\label{1.81-a}
\phi(x) \simeq \begin{cases}
-\displaystyle\frac{\sqrt{2}}{2a^2bx}, &x\to -\infty,\\[3mm]
a-\displaystyle\frac{2a}{b^2}(b^2+a^2)e^{-\mu x -2- 2a^2/b^2}, &x\to +\infty.
\end{cases}
\ee
Consequently, this kink is asymmetric due to the algebraic versus
exponential approach as $x \to\pm\infty$, respectively.
%Analogous expressions can be derived for the asymptotes at $\phi=-a,0$.
The kink's energy is
\begin{multline}\label{1.8u}
E_k = \frac{\sqrt{2}}{48} \lambda \Big[a\sqrt{b^2+a^2}(4a^4+4a^2 b^2+3b^4)\\
-3b^4(2a^2+b^2)\sinh^{-1}(a/b)\Big].
\end{multline}

\begin{figure}
\centerline{\includegraphics[width=0.5\textwidth]{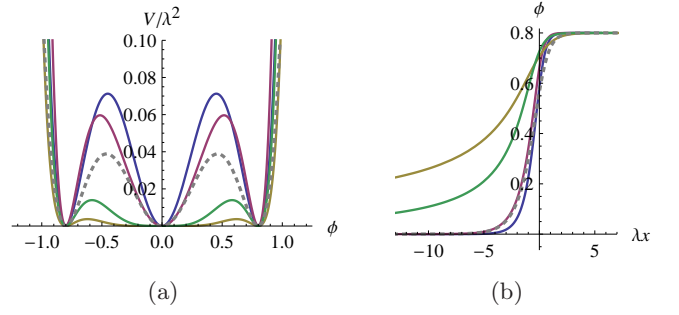}}
\hspace{1.65cm}(a)\hfill(b)\hspace{2cm}
\caption{(Color online.) $\phi^{10}$ field theory with three degenerate minima. (a) The potentials: (\ref{1.8g}) (first curve from top to bottom, blue online), (\ref{1.8c}) (second curve from top to bottom, red online), (\ref{1.15a}) (fourth curve from top to bottom, yellow online), (\ref{1.80}) (third curve from top to bottom, green online), and a representative $\phi^6$ potential $V(\phi) = \lambda^2\phi^2(\phi^2-a^2)^2$ (dotted). (b) The kink solutions connecting $0$ to $a$: (\ref{1.8b}) (fourth curve from top to bottom, blue online), (\ref{1.8e}) (third curve from top to bottom, red online), (\ref{1.15b}) (first curve from top to bottom, yellow online), (\ref{1.81}) (second curve from top to bottom, green online), and the corresponding $\phi^6$ kink $\phi(x) = a/\sqrt{1+e^{-2\sqrt{2}a^2\lambda x}}$ (dotted). In all panels, $a=4/5$, $b=1$ and $c=1$.}
\label{fig:phi10_3Degen_kinks}
\end{figure}

All four kinks from this subsection are illustrated
in Fig.~\ref{fig:phi10_3Degen_kinks}. Note that the plots for Cases III and IV are
distinct from those for Cases I and II in part due to the algebraic decay of the 
corresponding kink solutions as $\phi\to0$ [recall \eqref{1.15b-a} and \eqref{1.81-a}].

\subsection{Two Degenerate Minima}

There are three possible forms of the $\phi^{10}$ potential with two degenerate minima
at $\phi = \pm a$ for which kink solutions can be constructed.

\subsubsection{Case I}
Let
\be\label{1.15c}
V(\phi) = \lambda^2 (\phi^2-a^2)^2 (\phi^2+b^2)^3.
\ee
In this case,
\be
\begin{aligned}
\alpha_{8} &= a^2 - 3b^2,\\
\alpha_{6} &= 3b^4 - 6a^2b^2 + a^4,\\
\alpha_{4} &= b^2(6a^2b^2 - 3a^4 -b^4),\\
\alpha_{2} &= a^2b^4(3a^2 - 2b^2),\\
\alpha_{0} &= -a^4b^6,\\
\end{aligned}
\ee
It can be shown that, $\alpha_{8} < 0$ and $\alpha_0 < 0$,  while $\alpha_{6,4}>0$ and $\alpha_2 <0$
as long as $a < b\sqrt{3-\sqrt{6}} < \sqrt{3} a$.

The kink solution is given implicitly by
\begin{multline}\label{1.15m}
\mu x 
= \frac{2a\phi \sqrt{b^2+a^2}}{b^2\sqrt{b^2+\phi^2}} \\
+\sinh^{-1} \left [\frac{b^2+a\phi}{b(a-\phi)} \right]
-\sinh^{-1} \left [\frac{b^2-a\phi}{b(a+\phi)} \right],
\end{multline}
where $\mu = 2\sqrt{2} \lambda a (b^2+a^2)^{3/2}$.
From \eqref{1.15m}, the approach to the asymptotes
at $\phi = \pm a$ can be shown to be exponential:
\be
\phi(x) \simeq \begin{cases}
-a+\displaystyle\frac{2a}{b^2}(b^2+a^2)e^{\mu x + 2a^2/b^2}, &x\to -\infty,\\[3mm]
+a-\displaystyle\frac{2a}{b^2}(b^2+a^2)e^{-\mu x + 2a^2/b^2}, &x\to +\infty.
\end{cases}
\ee
Clearly, this kink is symmetric. The kink's energy is
\begin{multline}\label{1.15t}
E_k = \frac{\sqrt{2}}{24} \lambda
\Big[a\sqrt{b^2+a^2}(4a^4+16a^2b^2-3b^4)\\
+ 3b^4(6a^2+b^2)\sinh^{-1}(a/b) \Big].
\end{multline}

\subsubsection{Case II}
Let
\be\label{1.15l}
V(\phi) = \lambda^2 (\phi^2-a^2)^4 (\phi^2+b^2).
\ee
In this case,
\be
\begin{aligned}
\alpha_{8} &= 4a^2 - b^2,\\
\alpha_{6} &= a^2(6a^2 - 4b^2),\\
\alpha_{4} &= 2a^4(2a^2 - 3b^2),\\
\alpha_{2} &= a^6(a^2 - 4b^2),\\
\alpha_{0} &= -a^8b^2,\\
\end{aligned}
\ee
It can be shown that, $\alpha_{4,2,0} < 0$,  while $\alpha_{8,6}>0$
as long as $a \sqrt{6}/2 > b > a$.

The kink solution is given implicitly by
\begin{multline}\label{1.15k}
\mu x 
= \frac{2\phi a \sqrt{b^2+\phi^2}}{(a^2-\phi^2)\sqrt{b^2+a^2}}
+\left(\frac{2a^2+b^2}{b^2+a^2}\right)\\
\times\bigg\{ \sinh^{-1} 
\left[\frac{b^2+a\phi}{b(a-\phi)} \right]
- \sinh^{-1} \left[\frac{b^2-a\phi}{b(a+\phi)} \right]\bigg\},
\end{multline}
where $\mu = 4\sqrt{2}\lambda a^3 \sqrt{b^2+a^2}$.
From \eqref{1.15k}, the approach to the asymptotes
at $\phi = \pm a$ can be shown to be as $1/x$: 
\be
\phi(x) \simeq \begin{cases}
-a - \displaystyle\frac{a}{x}, &x\to -\infty,\\[3mm]
+a- \displaystyle\frac{a}{x}, &x\to +\infty.
\end{cases}
\ee
Clearly, this kink is symmetric. The kink's energy is
\begin{multline}\label{1.15u}
E_k = \frac{\sqrt{2}}{24} \lambda \Big[ 
a\sqrt{b^2+a^2}(8a^4-10a^2b^2-3b^4)\\
+ 3b^2(8a^4+4a^2 b^2+b^4)\sinh^{-1}(a/b) \Big].
\end{multline}

\subsubsection{Case III}
Let
\be\label{1.15g}
V(\phi) = \lambda^2 |\phi^2-a^2|^5,
\ee
In this case, $\alpha_{8,6,4,2,0} > 0$.
The kink solution is given implicitly by
\be\label{1.15h}
\mu x = \frac{\phi(3a^2-2\phi^2)}{(a^2-\phi^2)^{3/2}},
\ee
where $\mu = 3\sqrt{2} \lambda a^4$.
From \eqref{1.15h}, the approach to the asymptotes
at $\phi = \pm a$ can be shown to be algebraic:
\be
\phi(x) \simeq \begin{cases}
-a + \displaystyle\frac{a}{2(-\mu x)^{2/3}}, &x\to -\infty,\\[3mm]
+a- \displaystyle\frac{a}{2(\mu x)^{2/3}}, &x\to +\infty.
\end{cases}
\ee
Clearly, this kink is symmetric. The kink's energy is 
\be
E_k = \frac{5\sqrt{2} \pi }{16}\lambda a^6.
\ee

\begin{figure}
\centerline{\includegraphics[width=0.5\textwidth]{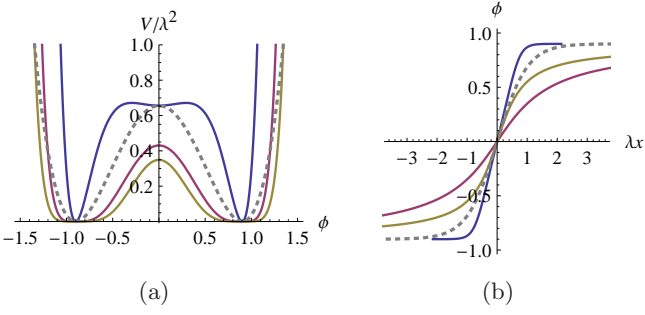}}
\hspace{1.65cm}(a)\hfill(b)\hspace{2cm}
\caption{(Color online.) $\phi^{10}$ field theory with two degenerate minima. (a) The potentials: (\ref{1.15c}) (top curve, blue online), (\ref{1.15l}) (middle curve, red online), (\ref{1.15g}) (bottom curve, yellow online), and a representative $\phi^4$ potential $V(\phi) = \lambda^2(\phi^2-a^2)^2$ (dotted). (b) The kink solutions connecting $-a$ to $a$: (\ref{1.15m}) (outer curve, blue online), (\ref{1.15k}) (inner curve, red online), (\ref{1.15h}) (middle curve, yellow online), and the corresponding $\phi^4$ kink $\phi(x) = a\tanh(\lambda x)$ (dotted). In all panels, $a=9/10$ and $b=1$.}
\label{fig:phi10_2Degen_kinks}
\end{figure}

All three kinks from this subsection are illustrated
in Fig.~\ref{fig:phi10_2Degen_kinks}.

\subsection{Phonons}
\label{sec:phi10-phonons}

The discussion from Section~\ref{sec:phi8-phonons} applies here as well. 
Table~\ref{table:phi10-phonos} summarizes the properties of the phonon
dispersion relation \eqref{eq:phi8-dispersion}
for the $\phi^{10}$ field theories with kink solutions studied above.
As was the case for the $\phi^8$ field theories considered above, 
there are once again potentials for which the RHS of the dispersion relation
vanishes; but, it cannot vanish in the other cases due to our assumption $b>a$.

\begin{table}
\caption{\label{table:phi10-phonos}Phonon modes of $\phi^{10}$ field theory. 
DM = degenerate minima. RHS = dispersion relation right-hand side.}
\begin{ruledtabular}
\begin{tabular}{lll}
potential, $V$ & equilibrium, $\phi_e$ &  RHS, $V''(\phi_e)$\\
\hline
5 DM, Eq.~\eqref{1.2} & $0$ & $2\lambda^2 a^4 b^4$\\
5 DM, Eq.~\eqref{1.2} & $\pm a$ & $8\lambda^2 a^4 (b^2-a^2)^2$\\
5 DM, Eq.~\eqref{1.2} & $\pm b$ & $8\lambda^2b^4 (b^2-a^2)^2$\\
\hline
4 DM, Eq.~\eqref{1.3} & $\pm a$ & $8\lambda^2a^2 (b^2-a^2)^2 (c^2+a^2)$\\
4 DM, Eq.~\eqref{1.3} & $\pm b$ & $8\lambda^2b^2 (b^2-a^2)^2 (c^2+b^2)$\\
\hline
3 DM, Eq.~\eqref{1.8g} & $0$ & $2\lambda^2ca^4$\\
3 DM, Eq.~\eqref{1.8g} & $\pm a$ & $8\lambda^2a^4 (a^4 -ba^2+c)$\\
\hline
3 DM, Eq.~\eqref{1.8c} & $0$ & $2\lambda^2a^4b^4$\\
3 DM, Eq.~\eqref{1.8c} & $\pm a$ & $8\lambda^2a^4 (b^2+a^2)^2$\\
\hline
3 DM, Eq.~\eqref{1.15a} & $0$ & $2\lambda^2a^4b^4$\\
3 DM, Eq.~\eqref{1.15a} & $\pm a$ & $8\lambda^2a^8$\\
\hline
3 DM, Eq.~\eqref{1.80} & $0$ & $2\lambda^2a^4b^4$\\
3 DM, Eq.~\eqref{1.80} & $\pm a$ & $8\lambda^2a^6(b^2+a^2)$\\
\hline
2 DM, Eq.~\eqref{1.15c} & $\pm a$ & $8\lambda^2 a^2(b^2+a^2)^3$\\
\hline
2 DM, Eq.~\eqref{1.15l} & $\pm a$ & $0$\\
\hline
2 DM, Eq.~\eqref{1.15g} & $\pm a$ & $0$\\
%\hline
%2 DM, Eq.~\eqref{1.10} & $\pm \hat{b} & 
\end{tabular}
\end{ruledtabular}
\end{table}

\subsection{Classical Free Energy Using the Transfer Matrix Technique}

Using the transfer matrix technique, it was shown by Scalapino et 
al.\ \cite{scal, ks} that, in the thermodynamic limit, the classical free energy
of a given field theory is essentially given by the ground state energy of
the Schr\"odinger-like equation whose potential is given by the field theory's
potential $V(\phi)$. Now, it is well known that while the ground state 
energy cannot be obtained analytically if the leading term of the potential
is of the form $\phi^{4n}$ with $n=1,2,\hdots$. On the other hand, if  the leading 
term in the potential is instead of the form $\phi^{4n+2}$, then it leads to a 
quasi-exactly solvable (QES) problem for which the eigenstates of the first few levels 
can be obtained analytically. For example, this has been demonstrated for the 
$\phi^{6}$ field theory in \cite{bk}. We would now like to show that there is a specific set 
of coefficients of our $\phi^{10}$ potential of form (\ref{1.1}) that leads to the 
classical free energy and probability distribution function (PDF) being obtainable 
analytically at a given temperature. 

In particular, the Schr\"odinger-like eigenvalue problem takes the form ($m=c=\hbar=1$):
\be\label{eq:schro}
-\frac{d^2\psi}{d\phi^2}+2V(\phi)\psi = 2E\psi,
\ee
with potential $V$ given by (\ref{1.1}). Then, it is easily shown that the
exact ground state energy eigenvalue and eigenfunction can be obtained exactly 
for some special cases of the coefficients $\alpha_i$ of the potential $V$.

First,
\be\label{eq:exact-state1}
\begin{aligned}
E_0 &= \frac{C}{2},\\
\psi_0(\phi) &= \exp\left[-\frac{\lambda \phi^6}{3 \sqrt{2}}+\frac{B\phi^4}{4}
-\frac{C\phi^2}{2}\right]
\end{aligned}
\ee
satisfy \eqref{eq:schro} provided $B$ and $C$ are related 
to $\lambda$, $\alpha_{8,6,4,2}$ via
\be
\begin{aligned}
\alpha_8 &= \frac{\sqrt{2} B}{\lambda},\\
\alpha_{6} &= \frac{B^2+2\sqrt{2} C\lambda}{2\lambda^2},\\
\alpha_{4} &= \frac{2BC+5\sqrt{2} \lambda}{2\lambda^2},\\
\alpha_2 &= \frac{C^2+3B}{2\lambda^2},\\
\alpha_0 &= 0.
\end{aligned}
\ee
This solution corresponds to a ground state.

Second,
\be\label{eq:exact-state2}
\begin{aligned}
E_1 &= \frac{C-F}{2},\\
\psi_1(\phi) &= (\phi^2+D)\exp\left[-\frac{\lambda \phi^6}{3 \sqrt{2}}+\frac{B\phi^4}{4}
-\frac{C\phi^2}{2}\right]
\end{aligned}
\ee
satisfy \eqref{eq:schro} provided $B$, $C$ and $D$ are related 
to $\lambda$, $\alpha_{8,6,4,2}$ via
\be
\begin{aligned}
\alpha_8 &= \frac{\sqrt{2} B}{\lambda},\\
\alpha_{6} &= \frac{B^2+2\sqrt{2} C\lambda}{2\lambda^2},\\
\alpha_{4} &= \frac{2BC+9\sqrt{2} \lambda}{2\lambda^2},\\
\alpha_2 &= \frac{C^2+3B+G}{2\lambda^2},\\
\alpha_0 &= 0,
\end{aligned}
\ee
where we have set $F=2/D$ and $G=-2(1+2CD)/D^2$ for convenience.
In addition, $D$ must satisfy the cubic equation
\be
2\sqrt{2}\lambda D^3 + 2BD^2 + 2CD + 1 = 0.
\ee
It is clear that as long as $B,C > 0$ (so that $\alpha_{8,6}>0$), then $D<0$, 
and the solution \eqref{eq:exact-state2} is for the second excited state (having two roots 
at $\phi = \pm\sqrt{D}$). %{\color{red}Actually, $B>0$ is necessary, while $C<0$ is allowed as long as $B>2\sqrt{2}\lambda$.}
On the other hand, if we allow $B < 0$, then $D > 0$ is possible for certain values of $C$, 
and the solution corresponds to another ground state. Note, however, that 
in that case $\alpha_8 < 0$.

Third,
\be\label{eq:exact-state3}
\begin{aligned}
E_2 &= \frac{C-G}{2},\\
\psi_2(\phi) &= (\phi^4+D\phi^2+J)\exp\left[-\frac{\lambda \phi^6}{3 \sqrt{2}}+\frac{B\phi^4}{4}
-\frac{C\phi^2}{2}\right]
\end{aligned}
\ee
satisfy \eqref{eq:schro} provided $B$, $C$, $D$ and $J$ are related 
to $\lambda$, $\alpha_{8,6,4,2}$ via
\be
\begin{aligned}
\alpha_8 &= \frac{\sqrt{2} B}{\lambda},\\
\alpha_{6} &= \frac{B^2+2\sqrt{2} C\lambda}{2\lambda^2},\\
\alpha_{4} &= \frac{2BC+13\sqrt{2} \lambda}{2\lambda^2},\\
\alpha_2 &= \frac{C^2+3B+H}{2\lambda^2},\\
\alpha_0 &= 0,
\end{aligned}
\ee
where we have set $G=2D/J$ and $H=(6/D-2D-G/2)G$ for convenience. %(6J-2CDJ-D^2)G
In addition, $D$ and $J$ must satisfy
\begin{align}
D^2 + 2(JC + \sqrt{2}\lambda J^2)D + 2J(2BJ - 3) &= 0,\\
D^3 + 2CJD^2 + (2BJ-7)DJ + 4J^2(\sqrt{2}\lambda J - C) &= 0.
\end{align}

The PDF for the classical field $\phi$ is just the square 
of the (normalized) ground state eigenfunctions given in
\eqref{eq:exact-state1}, \eqref{eq:exact-state2} or \eqref{eq:exact-state3}.

\section{$\phi^{12}$ Field Theory}

Finally, there are systems in which phase transitions are only captured by going to
the $\phi^{12}$ field theory (e.g., highly piezoelectric perovskite materials 
\cite{vc,piezo}). Depending on the form of the potential, it can have six, 
five, four, three or two degenerate minima, hence five, four, three, two 
or one kink solution(s) exist, respectively. In this section, we discuss these cases separately.
However, we do not provide a discussion of the various phases of the $\phi^{12}$ theory
because its complexity necessarily makes such a discussion quite lengthy.

\medskip
\subsection{Six Degenerate Minima}

Consider the potential
\be\label{3.1}
V(\phi) = \lambda^2 (\phi^2-a^2)^2 (\phi^2-b^2)^2 
(\phi^2 -c^2)^2,
\ee
where $c>b>a$ without  loss of generality.
This potential has six degenerate minima at $\phi= \pm a,\pm b,\pm c$
and, hence, five kink solutions exist.
Out of these five, only three are distinct due to the symmetry of the potential.

\subsubsection{Kink connecting $-a$ to $+a$}
In this case, the kink solution is given impicitly by
\begin{widetext}
\be\label{3.4}
e^{\mu x} = \left (\frac{a+\phi}{a-\phi} \right )^{(c^2-b^2)/a}
 \left (\frac{b-\phi}{b+\phi} \right )^{(c^2-a^2)/b}
\left (\frac{c+\phi}{c-\phi} \right )^{(b^2-a^2)/c},
\ee
where $\mu = 2\sqrt{2}\lambda (b^2-a^2)(c^2-b^2)(c^2-a^2)$.
From \eqref{3.4}, the approach to the asymptotes
at $\phi = \pm a$ can be shown to be exponential:
\be
\phi(x) \simeq \begin{cases}
-a +  2a \displaystyle \left[\left(\frac{b+a}{b-a}\right)^{(c^2-a^2)/b} \left(\frac{c-a}{c+a}\right)^{(b^2-a^2)/c} \right]^{-a/(c^2-b^2)} e^{\mu a x/(c^2-b^2)}, &x\to -\infty,\\[3mm]
+a - 2a \displaystyle \left[\left(\frac{b-a}{b+a}\right)^{(c^2-a^2)/b} \left(\frac{c+a}{c-a}\right)^{(b^2-a^2)/c} \right]^{a/(c^2-b^2)} e^{-\mu a x/(c^2-b^2)}, &x\to +\infty.
\end{cases}
\ee
Clearly, this kink is symmetric.
The corresponding kink energy is
\be\label{3.4h}
E_k^{(1)}= \frac{4\sqrt{2}}{105}  \lambda
a^3\left[3a^4-7(b^2+c^2)a^2 +35 b^2 c^2 \right].
\ee

\subsubsection{Kink connecting $a$ to $b$ (or $-b$ to $-a$)}
In this case, the kink solution is given implicitly by
\be\label{3.4d}
e^{\mu x} = \left (\frac{\phi-a}{\phi+a} \right )^{(c^2-b^2)/a}
 \left (\frac{b+\phi}{b-\phi} \right )^{(c^2-a^2)/b}
\left (\frac{c+\phi}{c-\phi} \right )^{(b^2-a^2)/c},
\ee
where $\mu$ is given below \eqref{3.4}.
From \eqref{3.4d}, the approach to the asymptotes
at $\phi = a,b$ can be shown to be exponential:
\be
\phi(x) \simeq \begin{cases}
a +  2 a \displaystyle \left[\left(\frac{b+a}{b-a}\right)^{(c^2-a^2)/b} \left(\frac{c+a}{c-a}\right)^{(b^2-a^2)/c} \right]^{-a/(c^2-b^2)} e^{\mu a x/(c^2-b^2)}, &x\to -\infty,\\[3mm]
b - 2b \displaystyle \left[\left(\frac{b-a}{b+a}\right)^{(c^2-b^2)/a} \left(\frac{c+b}{c-b}\right)^{(b^2-a^2)/c} \right]^{b/(c^2-a^2)} e^{-\mu b x/(c^2-a^2)}, &x\to +\infty.
\end{cases}
\ee
Due to the different growth rates, $\mu b/(c^2-a^2)$ versus $\mu a/(c^2-b^2)$ as $x\to\pm\infty$, respectively, this kink is asymmetric.
%Analogous expressions can be derived for the asymptotes at $\phi=-b,-a$.
The corresponding kink energy is
\be\label{3.4j}
E_k^{(2)} = \frac{2\sqrt{2} }{105} \lambda
(b-a)^3\big[7c^2(b^2+3ab+a^2)\\
-(3b^4+9b^3a+11b^2 a^2+9b a^3+3a^4) \big].
\ee

\subsubsection{Kink connecting $b$ to $c$ (or $-c$ to $-b$)}
In this case, the kink solution is given implicitly by
\be\label{3.4e}
e^{\mu x} = \left (\frac{\phi+a}{\phi-a} \right )^{(c^2-b^2)/a}
 \left (\frac{\phi-b}{\phi+b} \right )^{(c^2-a^2)/b}
\left (\frac{c+\phi}{c-\phi} \right )^{(b^2-a^2)/c},
\ee
where $\mu$ is given below \eqref{3.4}.
From \eqref{3.4e}, the approach to the asymptotes
at $\phi = b, c$ can be shown to be exponential:
\be
\phi(x) \simeq \begin{cases}
b +  2b \displaystyle \left[\left(\frac{b+a}{b-a}\right)^{(c^2-b^2)/a} \left(\frac{c+b}{c-b}\right)^{(b^2-a^2)/c} \right]^{-b/(c^2-a^2)} e^{\mu b x/(c^2-a^2)}, &x\to -\infty,\\[3mm]
c - 2c \displaystyle \left[\left(\frac{c+a}{c-a}\right)^{(c^2-b^2)/a} \left(\frac{c-b}{c+b}\right)^{(c^2-a^2)/b}  \right]^{c/(b^2-a^2)} e^{-\mu c x/(b^2-a^2)}, &x\to +\infty.
\end{cases}
\ee
\end{widetext}
Due to the different growth rates, $\mu c/(b^2-a^2)$ versus $\mu b/(c^2-a^2)$ as $x\to\pm\infty$, respectively, this kink is asymmetric.
%Analogous expressions can be derived for the asymptotes at $\phi=-c,-b$.

The corresponding kink energy is
\begin{multline}\label{3.4k}
E_k^{(3)} = \frac{2\sqrt{2}}{105} \lambda  (c-b)^3
\big[3c^4+9c^3b+11c^2 b^2+9c b^3\\
+3b^4-7a^2(c^2+3bc+b^2)\big].
\end{multline}

To the best of our knowledge, this is the first instance in which three
kink solutions exist for the same values of the potential's parameters. 
It would be of interest to determine values of the parameters $a$, $b$ and $c$ for 
which $E_k^{(1)}=E_k^{(2)}=E_k^{(3)}$.

\begin{figure}
\centerline{\includegraphics[width=0.5\textwidth]{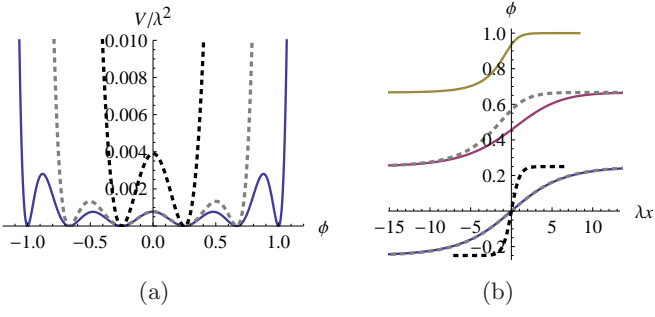}}
\hspace{1.65cm}(a)\hfill(b)\hspace{2cm}
\caption{(Color online.) $\phi^{12}$ field theory with six degenerate minima. (a) The potential (\ref{3.1}) (solid),  a representative $\phi^8$ potential (\ref{4}) (gray, dotted), and a representative $\phi^4$ potential $V(\phi) = \lambda^2(\phi^2-a^2)^2$ (black, dotted). (b) Kink solutions connecting $-a$ to $+a$ (\ref{3.4}) (bottom curve, blue online), $a$ to $b$ (\ref{3.4d}) (middle curve, red online), $b$ to $c$ (\ref{3.4e}) (top curve, yellow online), the corresponding $\phi^8$ kinks connecting $-a$ to $+a$ \eqref{4.2} and $a$ to $b$ \eqref{4.2b} (both gray, dotted), and the corresponding $\phi^4$ kink $\phi(x) = a\tanh(\lambda x)$ (black, dotted). In all panels, $a=1/4$, $b=2/3$ and $c=1$.}
\label{fig:phi12_6Degen_kinks}
\end{figure}

All three kinks from this subsection are illustrated
in Fig.~\ref{fig:phi12_6Degen_kinks} and compared to the 
kinks from the $\phi^8$ and $\phi^4$ field theories. Note that, while the
agreement between the $\phi^8$ kink connecting $-a$ to $+a$
and the corresponding $\phi^4$ one was quite good in Fig.~\ref{fig:phi8_TCI_kinks},
the agreement between the $\phi^{12}$ kink connecting $-a$ to $+a$
and the corresponding $\phi^4$ is not.
As in the previous examples, this is mainly due to the curvatures of the potentials near $\phi=0$ 
being quite different, hence the kinks having different widths.
The agreement between the $\phi^8$ and $\phi^{12}$ kinks connecting $-a$ to $+a$,
however, is so good that they are nearly indistinguishable for the chosen parameters.
On the other hand, the $\phi^{8}$ and $\phi^{12}$ kinks connecting $a$ to $b$ do not match as well.

\subsection{Five Degenerate Minima}

There are two possible forms of the $\phi^{12}$ potential 
with five degenerate minima for which we are
able to obtain the kink solutions. In this subsection, we discuss
these separately.

\subsubsection{Case I}
Consider the potential
\be\label{3.5}
V(\phi) = \lambda^2\phi^2 (\phi^2-a^2)^2 (\phi^2-b^2)^2 
(\phi^2 +c^2).
\ee
This potential has five degenerate minima at $\phi= 0,\pm a,\pm b$
and hence four kink solutions, two of which are distinct due to the symmetry of the potential.
As before, we take $b>a$ without any loss of generality. 
%In this case, 
%\be
%\begin{aligned}
%\alpha_{10} &= 2(b^2+a^2) - c^2,\\
%\alpha_{8} &= b^4  + 2 b^2 a^2 + a^4 - 2c^2(b^2+a^2),\\
%\alpha_{6} &= 2a^2b^2(b^2+a^2) - a^4c^2 - 4a^2b^2c^2 - b^4c^2,\\
%\alpha_{4} &= a^2b^2(a^2b^2 - 2a^2c^2-2b^2c^2),\\
%\alpha_{2} &= a^4b^4c^2,\\
%\alpha_{0} &= 0.
%\end{aligned}
%\ee

\paragraph{Kink connecting $0$ to $a$ (or $-a$ to $0$)}
This kink solution is given implicitly by
\begin{widetext}
\be\label{3.8}
e^{\mu x} = \left (\frac{\sqrt{c^2+\phi^2}-c}
{\sqrt{c^2+\phi^2}+c} \right ) \left (\frac{\sqrt{c^2+b^2}-\sqrt{c^2+\phi^2}}
{\sqrt{c^2+b^2}+\sqrt{c^2+\phi^2}} \right )^{a^2c/(b^2-a^2)\sqrt{c^2+b^2}}
 \left (\frac{\sqrt{c^2+a^2}+\sqrt{c^2+\phi^2}}
{\sqrt{c^2+a^2}-\sqrt{c^2+\phi^2}} \right )^{b^2c/(b^2-a^2)\sqrt{c^2+a^2}}, 
\ee
where $\mu = 2\sqrt{2}\lambda a^2 b^2 c$.
From \eqref{3.8},
it can be shown that the approach to the asymptotes at 
$\phi = 0,a$ is
\be\label{3.8-a}
\phi(x) \simeq \begin{cases}
2c \left[ \frac{a^2 +2 c(c + \sqrt{c^2+a^2})}{a^2}\right]^{-b^2c/2(b^2-a^2)\sqrt{c^2+a^2}}\left[ \frac{b^2 +2 c(c - \sqrt{c^2+b^2})}{b^2}\right]^{-a^2c/2(b^2-a^2)\sqrt{c^2+b^2}}e^{\mu x/2}, &x\to -\infty,\\[3mm]
a - \frac{2(c^2+a^2)}{a} \left(\frac{\sqrt{c^2+b^2}-\sqrt{c^2+a^2}}{\sqrt{c^2+b^2}+\sqrt{c^2+a^2}} \right)^{\frac{a^2\sqrt{c^2+a^2}}{b^2\sqrt{c^2+b^2}}} \left[ \frac{a^2 +2 c(c - \sqrt{c^2+a^2})}{a^2}\right]^{(b^2-a^2)\sqrt{c^2+a^2}/b^2c}e^{-\mu x (b^2-a^2)\sqrt{c^2+a^2}/b^2c}, &x\to +\infty.
\end{cases}
\ee
Consequently, this kink is asymmetric due to the different growth rates
as $\phi \to 0,a$.
%Analogous expressions can be derived for the asymptotes at $\phi=-a,0$.
The kink's energy is
\be\label{3.8j}
E_{k}^{(1)}= \frac{\sqrt{2}}{105}\lambda\left[2(c^2+a^2)^{3/2} (4c^4+7b^2-3a^2)-c^3(35a^2b^2+14a^2c^2+14b^2c^2+8c^4) \right].
\ee

\paragraph{Kink connecting $a$ to $b$ (or $-b$ to $-a$)}
This kink solution is given implicitly by
\be\label{3.8d}
e^{\mu x} = \left (\frac{\sqrt{c^2+\phi^2}-c}
{\sqrt{c^2+\phi^2}+c} \right ) \left (\frac{\sqrt{c^2+\phi^2}-\sqrt{c^2+a^2}}
{\sqrt{c^2+\phi^2}+\sqrt{c^2+a^2}} \right )^{b^2c/(b^2-a^2)\sqrt{c^2+a^2}}
\left (\frac{\sqrt{c^2+\phi^2}+\sqrt{c^2+b^2}}
{\sqrt{c^2+b^2}-\sqrt{c^2+\phi^2}} \right )^{a^2c/(b^2-a^2)\sqrt{c^2+b^2}}, 
\ee
where $\mu = 2\sqrt{2}\lambda a^2 b^2 c$.
From \eqref{3.8d},
it can be shown that the approach to the asymptotes at 
$\phi = a,b$ is
\be\label{3.8d-a}
\phi(x) \simeq \begin{cases}
a + \frac{2(c^2+a^2)}{a^2} \left(\frac{\sqrt{c^2+b^2}-\sqrt{c^2+a^2}}{\sqrt{c^2+b^2}+\sqrt{c^2+a^2}} \right)^{\frac{a^2\sqrt{c^2+a^2}}{b^2\sqrt{c^2+b^2}}} \left[ \frac{a^2 +2 c(c - \sqrt{c^2+a^2})}{a^2}\right]^{-(b^2-a^2)\sqrt{c^2+a^2}/b^2c}e^{\mu x (b^2-a^2)\sqrt{c^2+a^2}/b^2c}, &x\to -\infty,\\[3mm]
b - \frac{2(c^2+b^2)}{b^2} \left(\frac{\sqrt{c^2+b^2}-\sqrt{c^2+a^2}}{\sqrt{c^2+b^2}+\sqrt{c^2+a^2}} \right)^{\frac{b^2\sqrt{c^2+b^2}}{a^2\sqrt{c^2+a^2}}} \left[ \frac{b^2 +2 c(c - \sqrt{c^2+b^2})}{b^2}\right]^{(b^2-a^2)\sqrt{c^2+b^2}/a^2c}e^{- \mu x (b^2-a^2)\sqrt{c^2+b^2}/a^2c}, &x\to +\infty.
\end{cases}
\ee
\end{widetext}
Consequently, this kink is asymmetric due to the different growth rates
as $\phi \to a,b$.
%Analogous expressions can be derived for the asymptotes at $\phi=-b,-a$.
The kink's energy is
\begin{multline}\label{3.8jj}
E_{k}^{(2)}= \frac{2\sqrt{2}}{105} \lambda\bigg [ (a^2+c^2)^{3/2} 
(4c^2+7b^2-3a^2)\\
- (b^2+c^2)^{3/2} (4c^2-3b^2+7a^2) \bigg ].
\end{multline}

\subsubsection{Case II}
Now, consider the potential
\be\label{3.5d}
V(\phi) = \lambda^2 \phi^4 (\phi^2-a^2)^2 (\phi^2-b^2)^2 .
\ee
%This potential has five degenerate minima at $\phi= 0,\pm a,\pm b$
%and, hence, four kink solutions, two of which are distinct due to the symmetry of the potential. 
%As before, we take $b>a$ without any loss of generality.
In this case, 
\be
\begin{aligned}
\alpha_{10} &= 2(b^2+a^2),\\
\alpha_{8} &= b^4  + 2 b^2 a^2 + a^4,\\
\alpha_{6} &= 2a^2b^2(b^2+a^2),\\
\alpha_{4} &= b^4a^4,\\
\alpha_{2} &= \alpha_{0} = 0.
\end{aligned}
\ee
Clearly, $\alpha_{10,8,6,4} > 0$.

\paragraph{Kink connecting $0$ to $a$ (or $-a$ to $0$)}
This kink solution is given implicitly by
\be\label{3.8k}
\mu x = -\frac{2a(b^2-a^2)}{b^2 \phi}
+ \ln \left [\left(\frac{a+\phi}{a-\phi}\right)
\left(\frac{b-\phi}{b+\phi}\right)^{a^3/b^3} \right ],
\ee
where $\mu = 2\sqrt{2}\lambda a^3 (b^2-a^2)$.
From \eqref{3.8k},
it can be shown that the approach to the asymptotes at 
$\phi = 0,a$ is
\be\label{3.8k-a}
\phi(x) \simeq \begin{cases}
\displaystyle\frac{2a (b^2-a^2)}{b^2(-\mu x)}, &x\to -\infty,\\[3mm]
a - 2a \left(\frac{b-a}{b+a}\right)^{a^3/b^3}e^{2a^2/b^2 - \mu x - 2}, &x\to +\infty.
\end{cases}
\ee
Consequently, this kink is asymmetric due to the different growth types (algebraic 
versus exponential) as $\phi \to 0,a$, respectively.
%Analogous expressions can be derived for the asymptotes at $\phi=-a,0$.
The kink's energy is
\be\label{p1}
E_{k}^{(1)} = \frac{2\sqrt{2}}{105} \lambda a^5(7b^2-3a^2).
\ee

\paragraph{Kink connecting $a$ to $b$ (or $-b$ to $-a$)}
This kink is given implicitly by
\be\label{3.8f}
\mu x = \frac{2a(b^2-a^2)}{b^2 \phi}
+ \ln \left [\left(\frac{\phi-a}{\phi+a}\right) 
\left(\frac{b+\phi}{b-\phi}\right)^{a^3/b^3} \right] ,
\ee
where $\mu = 2\sqrt{2}\lambda a^3 (b^2-a^2)$.
From \eqref{3.8f},
it can be shown that the approach to the asymptotes at 
$\phi = a,b$ is
\begin{multline}\label{3.8f-a}
\phi(x) \simeq \\\begin{cases}
a + 2a\left(\frac{b-a}{b+a}\right)^{a^3/b^3}e^{\mu x - 2a(b^2-a^2)/b^a}, &x\to -\infty,\\[3mm]
b - 2b\left(\frac{b-a}{b+a}\right)^{b^3/a^3}e^{b^{3-b}(2b^2/a^2 - \mu x b^b/a^3 -2)}, &x\to +\infty.
\end{cases}
\end{multline}
Consequently, this kink is asymmetric due to the different growth rates
$\mu b^3/a^3$ versus  $\mu$ as $x \to \pm \infty$, repsectively.
%Analogous expressions can be derived for the asymptotes at $\phi=-b,-a$.
The kink's energy is
\be\label{p2}
E_{k}^{(2)} = \frac{2 \sqrt{2}}{105} \lambda (b-a)^{3}
\left[3b^4+9b^3 a+11b^2 a^2+9ba^3+3a^4 \right] .
\ee
Comparing the energies of the two kink solutions [(\ref{p1}) and (\ref{p2})], we find that 
$E_{k}^{(1)} \gtreqqless E_{k}^{(2)}$ if $b/a \lesseqqgtr \sqrt{7/3}$.

\begin{figure*}
\centerline{\includegraphics[width=0.8\textwidth]{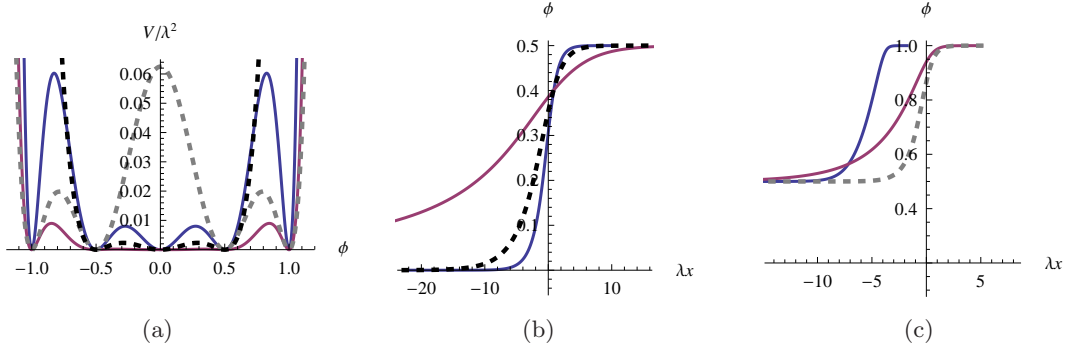}}
\hspace{3.6cm}(a)\hfill(b)\hfill(c)\hspace{3.8cm}
\caption{(Color online.) $\phi^{12}$ field theory with five degenerate minima. (a) The potentials: (\ref{3.5}) (top curve, blue online), (\ref{3.5d}) (bottom curve, red online), a representative $\phi^8$ potential with four degenerate minima (\ref{4}) (gray, dotted), and a representative $\phi^6$ potential with three degenerate minima $V(\phi) = \lambda^2 \phi^2 (\phi^2-a^2)^2$ (black, dotted). (b) The kink solutions connecting $0$ to $a$: (\ref{3.8}) (outer curve, blue online), (\ref{3.8k}) (inner curve, red online), and the corresponding $\phi^6$ kink $\phi(x) = a/\sqrt{1+e^{-2\sqrt{2}a^2\lambda x}}$ (black, dotted). 
(c) The kink solutions connecting $a$ to $b$: (\ref{3.8d}) (outer curve, blue online), (\ref{3.8f}) (inner curve, red online), and the corresponding $\phi^8$ kink (\ref{4.2b}) (gray, dotted).
In all panels, $a=1/2$, $b=1$ and $c=2$.}
\label{fig:phi12_5Degen_kinks}
\end{figure*}

All four kinks from this subsection are illustrated
in Fig.~\ref{fig:phi12_5Degen_kinks}. 

\subsection{Four Degenerate Minima} 
There are three possible forms of the $\phi^{12}$ potential 
with four degenerate minima for which we are
able to obtain the kink solutions. In this subsection, we discuss
these separately.

\subsubsection{Case I}
Consider the potential
\be\label{3.9}
V(\phi) = \lambda^2 (\phi^2-a^2)^2 (\phi^2-b^2)^2 (\phi^2+c^2)^2,
\ee
which has four degenerate minima at $\phi= \pm a,\pm b$ ($b>a$ as before) 
and, hence, three kink solutions, only two of which are distinct due to the 
symmetry of the potential.
%Here there are two distinct kink solutions and we discuss them one by one.

\paragraph{Kink connecting $-a$ to $+a$}
This kink solution is given implicitly by
\begin{widetext}
\be\label{3.10}
\mu x =  \frac{2a(b^2-a^2)}{c(c^2+b^2)} \tan^{-1}\left(\frac{\phi}{c}\right)
+ \frac{a(a^2+c^2)}{b(c^2+b^2)} \ln \left[\left(\frac{a+\phi}{a-\phi}\right) 
\left(\frac{b-\phi}{b+\phi}\right)\right],
\ee
where $\mu = 2\sqrt{2}\lambda a (c^2+a^2)(b^2-a^2)$.
From \eqref{3.10},
it can be shown that the approach to the asymptotes at 
$\phi = \pm a$ is exponential:
\be\label{3.10-a}
\phi(x) \simeq \begin{cases}
-a+2a\displaystyle\left(\frac{b-a}{b+a}\right)\exp\left[\frac{(b^2+c^2)\mu x b/a + 2(b^2-a^2)\tan^{-1}(a/c)b/c}{c^2+a^2}\right], &x\to -\infty,\\[4mm]
+a-2a\displaystyle\left(\frac{b-a}{b+a}\right)\exp\left[\frac{-(b^2+c^2)\mu x b/a + 2(b^2-a^2)\tan^{-1}(a/c)b/c}{c^2+a^2}\right], &x\to +\infty.
\end{cases}
\ee
Clearly, this kink is symmetric.
The kink's energy is
\be\label{3.10ff}
E_{k}^{(1)}=\frac{4\sqrt{2}}{105} \lambda a^3
[35b^2 c^2-7a^2 c^2 +7a^2 b^2 -3a^4].
\ee

\paragraph{Kink connecting $a$ to $b$ (or $-b$ to $-a$)}
This kink solution is given implicitly by
\be\label{3.10d}
\mu x =  -\frac{2a(b^2-a^2)}{c(c^2+b^2)} \tan^{-1}\left(\frac{\phi}{c}\right)
+ \frac{a(a^2+c^2)}{b(c^2+b^2)} \ln \left[\left(\frac{\phi-a}{\phi+a}\right) 
\left(\frac{b+\phi}{b-\phi}\right)\right],
\ee
where $\mu = 2\sqrt{2}\lambda a (c^2+a^2)(b^2-a^2)$.
From \eqref{3.10d},
it can be shown that the approach to the asymptotes at 
$\phi = a,b$ is exponential:
\be\label{3.10d-a}
\phi(x) \simeq \begin{cases}
a+2a\displaystyle\left(\frac{b-a}{b+a}\right)\exp\left[\frac{(b^2+c^2)\mu x b/a + 2(b^2-a^2)\tan^{-1}(a/c)b/c}{c^2+a^2}\right], &x\to -\infty,\\[4mm]
b-2b\displaystyle\left(\frac{b-a}{b+a}\right)\exp\left[\frac{-(b^2+c^2)\mu x b/a - 2(b^2-a^2)\tan^{-1}(b/c)b/c}{c^2+a^2}\right], &x\to +\infty.
\end{cases}
\ee
\end{widetext}
It can be shown that this kink is asymmetric.
%Analogous expressions can be derived for the asymptotes at $\phi=-b,-a$.
The kink's energy is
\begin{multline}\label{3.10p}
E_{k}^{(2)}=\frac{2\sqrt{2}}{105}\lambda (b-a)^{3}
\big[3b^4+9b^3 a+11b^2 a^2\\
+9ba^3+3a^4+7c^2(b^2+3ab+a^2)\big].
\end{multline}

\subsubsection{Case II}
Consider the potential
\be\label{3.9k}
V(\phi) = \lambda^2 (\phi^2-a^2)^4 (\phi^2-b^2)^2 .
\ee
%This potential has four degenerate minima at $\phi= \pm a,\pm b$
%and, hence, three kink solutions, two of which are distinct
%due to the symmetry of the potential.
In this case, 
\be
\begin{aligned}
\alpha_{10} &= 2(b^2+2a^2),\\
\alpha_{8} &= b^4  + 8 b^2 a^2 + 6a^4,\\
\alpha_{6} &= 4a^2(b^4 + 3a^2b^2 + a^4),\\
\alpha_{4} &= a^4(6b^4 + 8a^2b^2 +a^4),\\
\alpha_{2} &= 2a^6b^2(a^2 + 2b^2),\\
\alpha_{0} &= a^8b^4.
\end{aligned}
\ee
Clearly, $\alpha_{10,8,6,4,2,0} > 0$.
 
\paragraph{Kink connecting $-a$ and $+a$}
In this case, the kink solution is given implicitly by
\begin{multline}\label{3.10e}
\mu x =  \frac{b\phi(b^2-a^2)}{a^2(a^2-\phi^2)}  
+\ln\left(\frac{b+\phi}{b-\phi}\right)\\
-\frac{b(3a^2-b^2)}{2a^3} \ln\left(\frac{a+\phi}{a-\phi}\right),
\end{multline}
where $\mu = 2\sqrt{2}\lambda b(b^2-a^2)^2$.
From \eqref{3.10e},
it can be shown that the approach to the asymptotes at 
$\phi = \pm a$ is algebraic:
\be\label{3.10e-a}
\phi(x) \simeq \begin{cases}
-a-\displaystyle\frac{b(b^2-a^2)}{2a^2\mu x}, &x\to -\infty,\\[3mm]
+a-\displaystyle\frac{b(b^2-a^2)}{2a^2\mu x}, &x\to +\infty.
\end{cases}
\ee
Clearly, this kink is symmetric.
The kink's energy is
\be\label{3.10m}
E_{k}^{(1)}= \frac{16\sqrt{2} }{105} \lambda a^5(7b^2-a^2).
\ee

\paragraph{Kink connecting $a$ to $b$ (or $-b$ to $-a$)}
In this case, the kink solution is given implicitly by
\begin{multline}\label{3.10f}
\mu x =  -\frac{b\phi(b^2-a^2)}{a^2(\phi^2-a^2)}  
+\ln\left(\frac{b+\phi}{b-\phi}\right)\\
+\frac{b(3a^2-b^2)}{2a^3} \ln\left(\frac{\phi-a}{\phi+a}\right),
\end{multline}
where $\mu = 2\sqrt{2}\lambda b(b^2-a^2)^2$.
From \eqref{3.10f},
it can be shown that the approach to the asymptotes at 
$\phi = a,b$ is of mixed type:
\be\label{3.10f-a}
\phi(x) \simeq \begin{cases}
%a+2a\displaystyle\left(\frac{b-a}{b+a}\right)^{a^3/b^3}\exp\left[e^{\mu x} - 2a(b^2-a^2)/b^a\right], &x\to -\infty,\\[3mm]
a-\displaystyle\frac{b(b^2-a^2)}{2a^2\mu x}, &x\to -\infty,\\[3mm]
b-2b\left(\frac{b+a}{b-a}\right)^{(\kappa^{-2}-3)/(2\kappa)}e^{-\mu x -\kappa^{-2}}, &x\to +\infty,
\end{cases}
\ee
where $\kappa = a/b$.
Consequently, the kink is asymmetric due to the different growth types
as $x \to \pm\infty$.
%Analogous expressions can be derived for the asymptotes at $\phi=-b,-a$.
The kink's energy is
\be\label{3.10n}
E_{k}^{(2)}= \frac{2\sqrt{2}}{105}\lambda (b-a)^4(3b^3+12b^2 a+16ba^2+4a^3).
\ee

\begin{figure*}
\centerline{\includegraphics[width=0.8\textwidth]{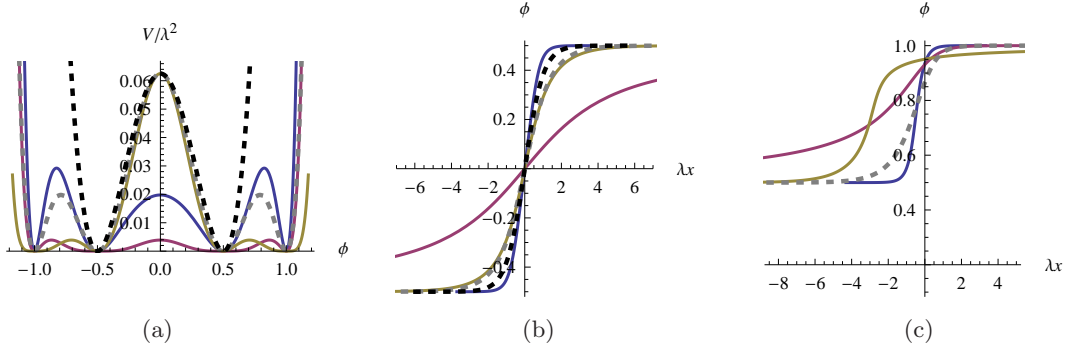}}
\hspace{3.6cm}(a)\hfill(b)\hfill(c)\hspace{3.8cm}
\caption{(Color online.) $\phi^{12}$ field theory with four degenerate minima. 
(a) The potentials: (\ref{3.9}) (second curve from top to bottom, blue online), (\ref{3.9k}) (third curve from top to bottom, red online), (\ref{3.9d}) (first curve from top to bottom, yellow online), a representative $\phi^8$ potential with four degenerate minima (\ref{4}) (gray, dotted), and a representative $\phi^4$ potential $V(\phi) = \lambda^2 (\phi^2-a^2)^2$ (black, dotted). 
(b) The kink solutions connecting $-a$ to $+a$: (\ref{3.10}) (outer curve, blue online), (\ref{3.10e}) (inner curve red, online), (\ref{3.10g}) (middle curve, yellow online), the corresponding $\phi^8$ kink (\ref{4.2}) (gray, dotted), and the corresponding $\phi^4$ kink $\phi(x) = a\tanh(\lambda x)$ (black, dotted). 
(c) The kink solutions connecting $a$ to $b$: (\ref{3.10d}) (outer curve, blue online), (\ref{3.10f}) (inner curve, red online), (\ref{3.10h}) (middle curve, yellow online), and the corresponding $\phi^8$ kink (\ref{4.2b}) (gray, dotted).
In all panels, $a=1/2$, $b=1$ and $c=3/4$.}
\label{fig:phi12_4Degen_kinks}
\end{figure*}

\subsubsection{Case III}
Consider the potential
\be\label{3.9d}
V(\phi) = \lambda^2 (\phi^2-a^2)^2 (\phi^2-b^2)^4 .
\ee
%This potential has four degenerate minima at $\phi= \pm a,\pm b$
%and, hence, three kink solutions, two of which are distinct.
In this case, 
\be
\begin{aligned}
\alpha_{10} &= 2(b^2+2a^2),\\
\alpha_{8} &= b^4  + 8 a^2 b^2 + 6 a^4,\\
\alpha_{6} &= 4a^2b^2(b^2+3a^2) + 4 a^6,\\
\alpha_{4} &= a^4(6b^4 + 8a^2b^2+a^4),\\
\alpha_{2} &= a^6b^2(4b^2+2a^2),\\
\alpha_{0} &= a^8b^4.
\end{aligned}
\ee
Clearly, $\alpha_{10,8,6,4,2,0} > 0$.
 
\paragraph{Kink connecting $-a$ and $+a$}
In this case, the kink solution is given implicitly by
\begin{multline}\label{3.10g}
\mu x = - \frac{a\phi(b^2-a^2)}{b^2(b^2-\phi^2)}  
+\ln\left(\frac{a+\phi}{a-\phi}\right)\\
-\frac{a(3b^2-a^2)}{2b^3} \ln\left(\frac{b+\phi}{b-\phi}\right),
\end{multline}
where $\mu = 2\sqrt{2}\lambda a(b^2-a^2)^2$.
From \eqref{3.10g},
it can be shown that the approach to the asymptotes at 
$\phi = \pm a$ is exponential:
\be\label{3.10g-a}
\phi(x) \simeq \begin{cases}
-a+2a\left(\frac{b+a}{b-a}\right)^{\kappa(\kappa^2-3)/2}e^{\mu x -\kappa^2}, &x\to -\infty,\\[3mm]
+a-2a\left(\frac{b+a}{b-a}\right)^{\kappa(\kappa^2-3)/2}e^{-\mu x -\kappa^2}, &x\to +\infty,
\end{cases}
\ee
where $\kappa = a/b$.
Clearly, this kink is symmetric.
The corresponding kink energy is
\be\label{pr63}
E_{k}^{(1)}=\frac{4\sqrt{2}}{105} \lambda a^3 (35 b^4-14a^2 b^2 +3a^4).
\ee

\paragraph{Kink connecting $a$ to $b$ (or $-b$ to $-a$)}
In this case, the kink solution is given implicitly by
\begin{multline}\label{3.10h}
\mu x =  \frac{a\phi(b^2-a^2)}{b^2(b^2-\phi^2)}  
+\ln\left(\frac{\phi-a}{\phi+a}\right)\\
+\frac{a(3b^2-a^2)}{2b^3} \ln\left(\frac{b+\phi}{b-\phi}\right),
\end{multline}
where $\mu = 2\sqrt{2}\lambda a(b^2-a^2)^2$.
From \eqref{3.10h},
it can be shown that the approach to the asymptotes at 
$\phi = a,b$ is of mixed type:
\be\label{3.10h-a}
\phi(x) \simeq \begin{cases}
%a+2a\displaystyle\left(\frac{b-a}{b+a}\right)^{a^3/b^3}\exp\left[e^{\mu x} - 2a(b^2-a^2)/b^a\right], &x\to -\infty,\\[3mm]
a-2a\left(\frac{b-a}{b+a}\right)^{\kappa(\kappa^2-3)/2}e^{\mu x -\kappa^2}, &x\to -\infty,\\[3mm]
b-\displaystyle\frac{a(b^2-a^2)}{2b^2\mu x}, &x\to +\infty.
\end{cases}
\ee
where $\kappa=a/b$.
Consequently, this kink is asymmetric due to the different growth types
as $x \to \pm \infty$.
%Analogous expressions can be derived for the asymptotes at $\phi=-b,-a$.
The kink's energy is
\be\label{p3}
E_{k}^{(2)}=\frac{2\sqrt{2}}{105} \lambda (b-a)^4(4 b^3+16b^2 a+12b a^2+3a^3).
\ee

All six kinks from this subsection are illustrated
in Fig.~\ref{fig:phi12_4Degen_kinks}.

\subsection{Three Degenerate Minima}

There are five possible forms of the $\phi^{12}$ potential with three degenerate minima 
for which kink solutions can be obtained analytically. In this subsection,
we discuss these cases separately.

\subsubsection{Case I}
Consider the potential
\be\label{3.11}
V(\phi) = \lambda^2 \phi^{8} (\phi^2-a^2)^2,
\ee
which has three degenerate minima at $\phi=0,\pm a$. In this case, 
$\alpha_{10,8}>0$, while $\alpha_{6,4,2,0}=0$. The kink solution,
which connects $0$ to $a$ (or $-a$ to $0$), as $x$ goes from 
$-\infty$ to $+\infty$, is given implicitly by
\be\label{3.12}
\mu x = -\frac {2a}{\phi}
-\frac{2a^3}{3\phi^3} + \ln \left(\frac{a+\phi}{a-\phi}\right),
\ee 
where $\mu = 2\sqrt{2} \lambda a^5$.
From \eqref{3.12},
it can be shown that the approach to the asymptotes at 
$\phi = 0,a$ is of mixed type:
\be
\phi(x) \simeq \begin{cases}
\displaystyle\frac{2^{1/3}a}{(-3\mu x)^{1/3}}, &x\to -\infty,\\[3mm]
a-2ae^{-\mu x - 8/3}, &x\to +\infty.
\end{cases}
\ee
Consequently, this kink is asymmetric due to the different growth types
as $x\to\pm\infty$.
%Analogous expressions can be derived for the asymptotes at $\phi=-a,0$.
The kink's energy is
\be\label{3.12j}
E_k = \frac{2\sqrt{2}}{35} \lambda a^7.
\ee

\subsubsection{Case II}
Consider the potential
\be\label{3.13}
V(\phi) = \lambda^2 \phi^{4} (\phi^2-a^2)^4.
\ee
%which has three degenerate minima at $\phi=0,\pm a$. 
In this case, 
$\alpha_{10,8,6,4}>0$, while $\alpha_{2,0}=0$. The kink solution,
which connects $0$ to $a$ (or $-a$ to $0$), as $x$ goes from 
$-\infty$ to $+\infty$, is given implicitly by
\be\label{3.14}
\mu x = \frac{2a(3\phi^2-2a^2)}{3\phi(a^2-\phi^2)}
+\ln \left(\frac{a+\phi}{a-\phi}\right),
\ee
where $\mu = (4/3)\sqrt{2} \lambda a^5$.
From \eqref{3.14},
it can be shown that the approach to the asymptotes at 
$\phi = 0,a$ is algebraic:
\be\label{3.14-a}
\phi(x) \simeq \begin{cases}
\displaystyle -\frac{a}{\sqrt{2}  a^5 \lambda x}, &x\to -\infty,\\[3mm]
a-\displaystyle \frac{a}{4\sqrt{2}  a^5 \lambda x}, &x\to +\infty.
\end{cases}
\ee
Consequently, this kink is asymmetric due to the different growth rates
as $x \to \pm \infty$.
%Analogous expressions can be derived for the asymptotes at $\phi=-a,0$.
The kink's energy is
\be\label{3.14j}
E_k = \frac{8\sqrt{2}}{105} \lambda a^7.
\ee

\subsubsection{Case III}
Consider the potential
\be\label{3.15}
V(\phi) = \lambda^2 \phi^{4} (\phi^2-a^2)^2 (\phi^2+b^2)^2.
\ee
%which has three degenerate minima at $\phi=0,\pm a$.
In this case,
\be
\begin{aligned}
\alpha_{10} &= 2(b^2 - a^2),\\
\alpha_{8} &= b^4 - 4a^2b^2 + a^4,\\
\alpha_{6} &= 2a^2b^2(b^2 - a^2),\\
\alpha_{4} &= a^4b^4,\\
\alpha_{2} &= \alpha_{0} = 0.
\end{aligned}
\ee
It can be shown that $\alpha_{10,8,6,4}>0$ as long as $b\sqrt{2-\sqrt{3}} > a$.

The kink solution,
which connects $0$ to $a$ (or $-a$ to $0$), as $x$ goes from 
$-\infty$ to $+\infty$, is given implicitly by
\be\label{3.16}
\mu x = 
-\frac{2a(b^2+a^2)}{b^2\phi} -\frac{2a^3}{b^3} \tan^{-1}\left(\frac{\phi}{b}\right)
+\ln \left(\frac{a+\phi}{a-\phi}\right),
\ee
where $\mu = 2\sqrt{2} \lambda a^3 (b^2+a^2)$.
From \eqref{3.16},
it can be shown that the approach to the asymptotes at 
$\phi = 0,a$ is of mixed type:
\be
\phi(x) \simeq \begin{cases}
\displaystyle -\frac{1}{\sqrt{2} b^2a^2\lambda   x}, &x\to -\infty,\\[3mm]
a-2ae^{-\mu x -2(1+\kappa^2+\kappa^3\tan^{-1}\kappa)}, &x\to +\infty,
\end{cases}
\ee
where $\kappa = a/b$.
Consequently, this kink is asymmetric due to the different growth types
as $x \to \pm\infty$.
%Analogous expressions can be derived for the asymptotes at $\phi=-a,0$.
The kink's energy is
\be\label{3.16j}
E_k = \frac{2\sqrt{2}}{105} \lambda a^5(7b^2+3a^2).
\ee

\subsubsection{Case IV}
Consider the potential
\be\label{3.17}
V(\phi) = \lambda^2 \phi^{2} (\phi^2-a^2)^2 (\phi^2+b^2)^{3}.
\ee
%which has three degenerate minima at $\phi=0,\pm a$. 
In this case,
\be
\begin{aligned}
\alpha_{10} &= 2a^2 - 3b^2,\\
\alpha_{8} &= 3b^4 - 6a^2b^2 + a^4,\\
\alpha_{6} &= b^2(6 a^2b^2 - 3a^4 -b^4),\\
\alpha_{4} &= a^2b^4(3a^2-2b^2),\\
\alpha_{2} &= -a^4b^6,\\
\alpha_{0} &= 0.
\end{aligned}
\ee
It can be shown that $\alpha_{10,2} < 0$, while $\alpha_{8,6}>0$ and $\alpha_{4} <0$ as long as 
$a < b\sqrt{3-\sqrt{6}} < \sqrt{3} a$.

The kink solution,
which connects $0$ to $a$ (or $-a$ to $0$) as $x$ goes from 
$-\infty$ to $+\infty$, is given implicitly by
\begin{widetext}
\be\label{3.18}
\mu x 
= \frac{2a^2\sqrt{b^2+a^2}}{b^2\sqrt{b^2+\phi^2}}
 + \frac{(b^2+a^2)^{3/2}}{b^3} \ln \left [\left(\frac{\sqrt{b^2+a^2}+\sqrt{b^2+\phi^2}}
{\sqrt{b^2+a^2}-\sqrt{b^2+\phi^2}}\right)
\left (\frac{\sqrt{b^2+\phi^2}-b}{\sqrt{b^2+\phi^2}+b} \right ) \right ],
\ee
where $\mu = \sqrt{2} \lambda (b^2+a^2)^{3/2}$.
From \eqref{3.18},
it can be shown that the approach to the asymptotes at 
$\phi = 0,a$ is exponential:
\be\label{3.18-a}
\phi(x) \simeq \begin{cases}
\displaystyle\frac{2 a b}{\sqrt{a^2+2 b \left(b+\sqrt{b^2+a^2}\right)}} \exp\left[-\frac{a^2}{b^2+a^2}+\frac{b^3 \mu x }{2 \left(b^2+a^2\right)^{3/2}}\right], &x\to -\infty,\\[3mm]
\displaystyle \frac{a}{\left(b+\sqrt{b^2+a^2}\right)^2} \left\{a^2+2 b \left(b+\sqrt{b^2+a^2}\right)-2 \left(b^2+a^2\right) \exp\left[\frac{2 a^2 b-b^3 \mu x}{\left(b^2+a^2\right)^{3/2}}\right]\right\}, &x\to +\infty.
\end{cases}
\ee
\end{widetext}
Consequently, this kink is asymmetric due to the different growth rates
as $x \to \pm\infty$.
%Analogous expressions can be derived for the asymptotes at $\phi=-a,0$.
The kink's energy is
\be\label{3.18j}
%E_k = \frac{2\sqrt{2}}{35} \lambda (b^2+a^2)^{7/2}.
E_k = \frac{\sqrt{2}}{35} \lambda \left[ 2(b^2+a^2)^{7/2}
 - b^5(7a^2+b^2) \right].
\ee

\subsubsection{Case V}
Consider the potential
\be\label{3.17r}
V(\phi) = \lambda^2 \phi^{6} (\phi^2-a^2)^2 (\phi^2+b^2),
\ee
which has three degenerate minima at $\phi=0,\pm a$. 
In this case,
\be
\begin{aligned}
\alpha_{10} &= b^2 - 2a^2,\\
\alpha_{8} &= a^4 - 2a^2b^2,\\
\alpha_{6} &= -a^4b^2,\\
\alpha_{4} &= \alpha_{2} = \alpha_{0} = 0.
\end{aligned}
\ee
It can be shown that $\alpha_{10}>0$ as long as $b>\sqrt{2}a$, while $\alpha_{8,6} < 0$.

The kink solution,
which connects $0$ to $a$ (or $-a$ to $0$), as $x$ goes from 
$-\infty$ to $+\infty$, is given implicitly by
\begin{widetext}
\be\label{3.18r}
\mu x 
= -\frac{a^2\sqrt{b^2+a^2}\sqrt{b^2+\phi^2}}{b^2 \phi^2}
+ \frac{(2b^2-a^2)(b^2+a^2)^{1/2}}{b^3} \ln \left [\left(\frac{\sqrt{b^2+\phi^2}+\sqrt{b^2+a^2}}
{\sqrt{b^2+a^2}-\sqrt{b^2+\phi^2}}\right)
\left (\frac{\sqrt{b^2+\phi^2}-b}{
\sqrt{b^2+\phi^2} + b} \right ) \right ],
\ee
where $\mu = 2\sqrt{2} \lambda a^4 (b^2+a^2)^{1/2}$.
From \eqref{3.18r},
it can be shown that the approach to the asymptotes at 
$\phi = 0,a$ is of mixed type:
\be
\phi(x) \simeq \begin{cases}
\displaystyle\frac{a(b^2+a^2)^{1/4}}{\sqrt{-b\mu x}}, &x\to -\infty,\\[3mm]
a -a^{-3}\left[2a^4+6a^2b^2+4b^4-4b(b^2+a^2)^{3/2}\right] e^{-\mu x - 1 - a^2/b^2}, &x\to +\infty.
\end{cases}
\ee
\end{widetext}
Consequently, this kink is asymmetric due to the different growth types
as $x\to \pm\infty$.
%Analogous expressions can be derived for the asymptotes at $\phi=-a,0$.
The kink's energy is
\be\label{3.18l}
E_k = \frac{2\sqrt{2}}{105}\lambda \left[(4b^2+7a^2)b^5 
-(4b^2-3a^2)(b^2+a^2)^{5/2}\right].
\ee

\begin{figure}[h]
\centerline{\includegraphics[width=0.5\textwidth]{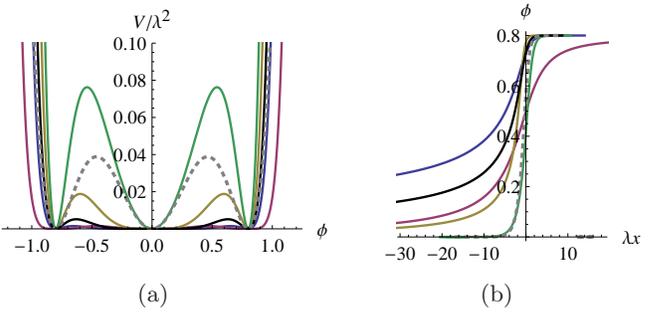}}
\hspace{1.65cm}(a)\hfill(b)\hspace{2cm}
\caption{(Color online.) $\phi^{12}$ field theory with three degenerate minima. (a) The potentials: (\ref{3.11}) (fourth curve from top to bottom, blue online), (\ref{3.13}) (fifth curve from top to bottom, red online), (\ref{3.15}) (second curve from top to bottom, yellow online), (\ref{3.17}) (first curve from top to bottom, green online), (\ref{3.17r}) (third curve from top to bottom, black online), and a representative $\phi^6$ potential $V(\phi) = \lambda^2\phi^2(\phi^2-a^2)^2$ (gray, dotted). (b) The kink solutions connecting $0$ to $a$: (\ref{3.12}) (first curve from top to bottom, blue online), (\ref{3.14}) (third curve from top to bottom, red online), (\ref{3.16}) (fourth curve from top to bottom, yellow online), (\ref{3.18}) (fifth curve from top to bottom, green online), (\ref{3.18r}) (second curve from top to bottom, black online), and the corresponding $\phi^6$ kink $\phi(x) = a/\sqrt{1+e^{-2\sqrt{2}a^2\lambda x}}$ (gray, dotted). In all panels, $a=4/5$ and $b=1$.}
\label{fig:phi12_3Degen_kinks}
\end{figure}

All five kinks from this subsection are illustrated
in Fig.~\ref{fig:phi12_3Degen_kinks}. Note that the plots for Cases II and IV 
are distinct from those for Cases I, III and V in part due to pure algebraic and
pure exponential versus mixed type, respectively, decay of the 
corresponding kinks' tails as $\phi\to0,a$ [recall \eqref{3.14-a} and \eqref{3.18-a}].

\subsection{Two Degenerate Minima}

There are three possible forms of the $\phi^{12}$ potential with
two degenerate minima for which we can 
obtain a kink solution that connects $\phi=-a$ to $\phi=+a$,
as $x$ goes from $-\infty$ to $+\infty$. We discuss these separately.

\subsubsection{Case I}
Consider the potential
\be\label{3.19}
V(\phi)=\lambda^2 (\phi^2-a^2)^2 (\phi^2+b^2)^{4}.
\ee
In this case,
\be
\begin{aligned}
\alpha_{10} &= 2(2b^2 - a^2),\\
\alpha_{8} &= 6b^4 - 8a^2b^2 + a^4,\\
\alpha_{6} &= 4b^2(b^2 + ab - a^2)(a^2 + ab - b^2),\\
\alpha_{4} &= b^4(b^4 - 8a^2b^2 + 6a^4),\\
\alpha_{2} &= 2a^2b^6(b^2-2a^2),\\
\alpha_{0} &= a^4b^8.
\end{aligned}
\ee
It can be shown that $\alpha_{10,0}>0$, while $\alpha_{8,6,2} > 0$ and $\alpha_4 < 0$ as long as
$2a/(\sqrt{5}-1) > b > \sqrt{2} a$.
%> a/\sqrt{4-\sqrt{10}}$

The kink solution is given implicitly by 
\begin{multline}\label{3.20}
\mu x = \frac{a(b^2+a^2)\phi}{b^2(b^2+\phi^2)}
+\frac{a(a^2+3b^2)}{b^3} \tan^{-1} \left(\frac{\phi}{b} \right)\\
+\ln \left(\frac{a+\phi}{a-\phi}\right),
\end{multline}
where $\mu = 2\sqrt{2}\lambda a(b^2+a^2)^{2}$.
From \eqref{3.20}, the approach to the asymptotes
at $\phi = \pm a$ can be shown to be exponential:
\be
\phi(x) \simeq \begin{cases}
-a +  2a e^{\mu x +\kappa^2+(\kappa^2+3)\kappa\tan^{-1}\kappa}, &x\to -\infty,\\[3mm]
+a - 2a e^{-\mu x +\kappa^2+(\kappa^2+3)\kappa\tan^{-1}\kappa}, &x\to +\infty,
\end{cases}
\ee
where $\kappa = a/b$.
Clearly, this kink is symmetric.
The kink's energy is
\be\label{3.20j}
E_k = \frac{4\sqrt{2}}{105} \lambda a^3(35 b^4+14 a^2 b^2 + 3a^4).
\ee

\subsubsection{Case II}
Consider the potential
\be\label{3.23}
V(\phi)=\lambda^2 (\phi^2-a^2)^4 (\phi^2+b^2)^2.
\ee
In this case,
\be
\begin{aligned}
\alpha_{10} &= 2(2a^2 - b^2),\\
\alpha_{8} &= b^4 - 8a^2b^2 +  6a^4,\\
\alpha_{6} &= 4a^2(a^2 - ab - b^2)(a^2 + ab - b^2),\\
\alpha_{4} &= a^4(6b^4 - 8a^2b^2 + a^4),\\
\alpha_{2} &= 2a^6b^2(2b^2-a^2),\\
\alpha_{0} &= a^8b^4.
\end{aligned}
\ee
It can be shown that $\alpha_{10}<0$ and $\alpha_{8,6,4,2}>0$ as long as 
$b\sqrt{4-\sqrt{10}} > \sqrt{6} a$, while $\alpha_0>0$.

The kink solution is given implicitly by
\begin{multline}\label{3.24}
\mu x 
= \frac{2a(b^2+a^2)\phi}{(3a^2+b^2)(a^2-\phi^2)}
+\frac{4a^3}{b(3a^2+b^2)}\tan^{-1}\left(\frac{\phi}{b}\right)\\
+\ln \left(\frac{a+\phi}{a-\phi}\right),
\end{multline}
where $\mu = 4\sqrt{2}\lambda (b^2+a^2)^2/(3a^2+b^2)$.
From \eqref{3.24}, the approach to the asymptotes
at $\phi = \pm a$ can be shown to be algebraic:
\be
\phi(x) \simeq \begin{cases}
-a - \displaystyle \frac{a}{4\sqrt{2}(b^2+a^2)\lambda x}, &x\to -\infty,\\[3mm]
+a - \displaystyle \frac{a}{4\sqrt{2}(b^2+a^2)\lambda x}, &x\to +\infty.
\end{cases}
\ee
Clearly, this kink is symmetric.
The kink's energy is
\be\label{3.24j}
E_k = \frac{16\sqrt{2}}{105}\lambda a^5(7 b^2+ a^2).
\ee

\subsubsection{Case III} 
Consider the potential
\be\label{3.21}
V(\phi)=\lambda^2 (\phi^2-a^2)^6.
\ee
In this case, $\alpha_{10,8,6,4,2,0} > 0$.
The kink solution is given implicitly by
\be\label{3.22}
\mu x = \frac{a(7a^2-3\phi^2)\phi}{3 (a^2-\phi^2)^2}
+\ln \left(\frac{a+\phi}{a-\phi}\right),
\ee
where $\mu = (16/3)\sqrt{2}a^5 \lambda$.
From \eqref{3.22}, the approach to the asymptotes
at $\phi = \pm a$ can be shown to be algebraic:
\be
\phi(x) \simeq \begin{cases}
-a +  \displaystyle\frac{a}{\sqrt{-3 \mu x}}, &x\to -\infty,\\[3mm]
+a - \displaystyle\frac{a}{\sqrt{3\mu x}}, &x\to +\infty.
\end{cases}
\ee
Clearly, this kink is symmetric.
The kink's energy is
\be\label{3.22j}
E_k = \frac{32\sqrt{2}}{35}\lambda a^{7}.
\ee

\begin{figure}
\centerline{\includegraphics[width=0.5\textwidth]{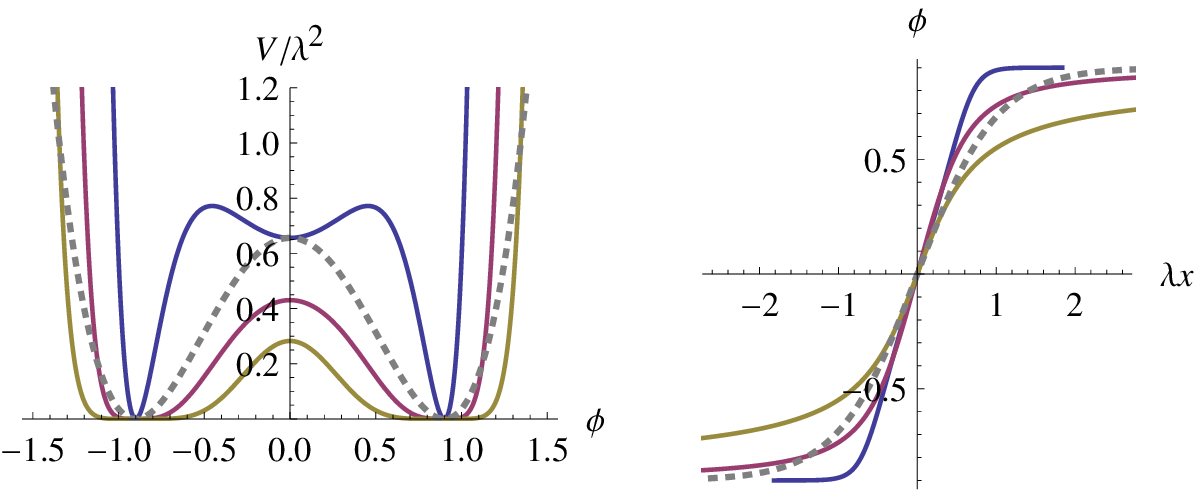}}
\hspace{1.65cm}(a)\hfill(b)\hspace{2cm}
\caption{(Color online.) $\phi^{12}$ field theory with two degenerate minima. (a) The potentials: (\ref{3.19}) (top curve, blue online), (\ref{3.23}) (middle curve, red online), (\ref{3.21}) (bottom curve, yellow online), and a representative $\phi^4$ potential $V(\phi) = \lambda^2(\phi^2-a^2)^2$ (dotted). (b) The kink solutions connecting $-a$ to $+a$: (\ref{3.20}) (outer curve, blue online), (\ref{3.24}) (middle curve, red online), (\ref{3.22}) (inner curve, yellow online), and the corresponding $\phi^4$ kink $\phi(x) = a\tanh(\lambda x)$ (dotted). In all panels, $a=9/10$ and $b=1$.}
\label{fig:phi12_2Degen_kinks}
\end{figure}

All three kinks from this subsection are illustrated
in Fig.~\ref{fig:phi12_2Degen_kinks}.

\subsection{Phonons}

The discussion from Section~\ref{sec:phi8-phonons} applies here as well. 
Table~\ref{table:phi12-phonos} summarizes the properties of the phonon
dispersion relation \eqref{eq:phi8-dispersion}
for the $\phi^{12}$ field theories with kink solutions studied above.
As was the case for the $\phi^8$ and $\phi^{10}$ field theories considered above, 
there are once again potentials for which the RHS of the dispersion relation
vanishes; but, it cannot vanish in the other cases due to our assumption $c>b>a$.

\begin{table}[h]
\caption{\label{table:phi12-phonos}Phonon modes of $\phi^{12}$ field theory. 
DM = degenerate minima. RHS = dispersion relation right-hand side.}
\begin{ruledtabular}
\begin{tabular}{lll}
potential, $V$ & equilibrium, $\phi_e$ &  RHS, $V''(\phi_e)$\\
\hline
6 DM, Eq.~\eqref{3.1} & $\pm a$ & $8\lambda^2 a^2 (b^2-a^2)^2 (c^2-a^2)^2$\\
6 DM, Eq.~\eqref{3.1} & $\pm b$ & $8\lambda^2 b^2 (b^2-a^2)^2 (c^2-b^2)^2$\\
6 DM, Eq.~\eqref{3.1} & $\pm c$ & $8\lambda^2 c^2 (c^2-a^2)^2 (c^2-b^2)^2$\\
\hline
5 DM, Eq.~\eqref{3.5} & $0$ & $2\lambda^2 a^4 b^4 c^2$\\
5 DM, Eq.~\eqref{3.5} & $\pm a$ & $8\lambda^2 a^4 (b^2-a^2)^2 (c^2+a^2)$\\
5 DM, Eq.~\eqref{3.5} & $\pm b$ & $8\lambda^2 b^4 (b^2-a^2)^2 (c^2+b^2)$\\
\hline
5 DM, Eq.~\eqref{3.5d} & $0$ & $0$\\
5 DM, Eq.~\eqref{3.5d} & $\pm a$ & $8\lambda^2 a^6 (b^2-a^2)^2$\\
5 DM, Eq.~\eqref{3.5d} & $\pm b$ & $8\lambda^2 b^6 (b^2-a^2)^4$\\
\hline
4 DM, Eq.~\eqref{3.9} & $\pm a$ & $8\lambda^2 a^2 (b^2-a^2)^2 (c^2+a^2)^2$\\
4 DM, Eq.~\eqref{3.9} & $\pm b$ & $8\lambda^2 b^2 (b^2-a^2)^2 (c^2+b^2)^2$\\
\hline
4 DM, Eq.~\eqref{3.9k} & $\pm a$ & $0$\\
4 DM, Eq.~\eqref{3.9k} & $\pm b$ & $8\lambda^2 b^2(b^2-a^2)^4$\\
\hline
4 DM, Eq.~\eqref{3.9d} & $\pm a$ & $8\lambda^2 a^2(b^2-a^2)^4$\\
4 DM, Eq.~\eqref{3.9d} & $\pm b$ & $0$\\
\hline
3 DM, Eq.~\eqref{3.11} & $0$ & $0$\\
3 DM, Eq.~\eqref{3.11} & $\pm a$ & $8\lambda^2a^{10}$\\
\hline
3 DM, Eq.~\eqref{3.13} & $0$ & $0$\\
3 DM, Eq.~\eqref{3.13} & $\pm a$ & $0$\\
\hline
3 DM, Eq.~\eqref{3.15} & $0$ & $0$\\
3 DM, Eq.~\eqref{3.15} & $\pm a$ & $8\lambda^2 a^6(b^2+a^2)^2$\\
\hline
3 DM, Eq.~\eqref{3.17} & $0$ & $2\lambda^2a^4b^6$\\
3 DM, Eq.~\eqref{3.17} & $\pm a$ & $8\lambda^2a^4(b^2+a^2)^3$\\
\hline
3 DM, Eq.~\eqref{3.17r} & $0$ & $0$\\
3 DM, Eq.~\eqref{3.17r} & $\pm a$ & $8\lambda^2a^8(b^2+a^2)$\\
\hline
2 DM, Eq.~\eqref{3.19} & $\pm a$ & $8\lambda^2 a^2 (b^2+a^2)^4$\\
\hline
2 DM, Eq.~\eqref{3.23} & $\pm a$ & $0$\\
\hline
2 DM, Eq.~\eqref{3.21} & $\pm a$ & $0$\\
%\hline
%2 DM, Eq.~\eqref{1.10} & $\pm \hat{b} & 
\end{tabular}
\end{ruledtabular}
\end{table}

\section{Limiting Behaviors as $n\to\infty$}

As the degree of the even polynomial field theories considered herein becomes large,
there are two limiting cases to be considered. The potentials have the general form
\be
V_{2m}(\phi) = \lambda^2 \sum_{i=0}^{m} (-1)^{m-i}\alpha_{2i}\phi^{2i},
\ee
where for $m=2n$ (even) we obtain the $\phi^4$, $\phi^8$, $\phi^{12}$, etc.\ field theories,
whereas for $m=2n+1$ (odd) we obtain the $\phi^6$, $\phi^{10}$, etc.\ field theories.

Now, there are two paths to obtaining the limiting field theory as $m\to\infty$. First, for 
$m=2n$ (even),  we choose $\alpha_0=2$ and $\alpha_{2i} = 1/(2i)!$, then
\be
\lim_{n\to\infty}V_{4n}(\phi) = \lambda^2(1+\cos\phi), 
\ee
which satisfies both $\min_\phi V(\phi) = 0$, $\alpha_{2i}>0$ for all $i$, and the coefficient of $\phi^{4n}$ is $\alpha_{4n} > 0$ as needed to ensure $V_{4n}(\phi) \to +\infty$ as $|\phi| \to \infty$. For these theories, the maximum
number of degenerate minima is even and, hence, there is no degenerate minimum at $\phi=0$, 
\emph{unlike} the sine-Gordon theory.

Second, for $m=2n+1$ (odd), we choose $\alpha_0=0$ and $\alpha_{2i} = 1/(2i)!$, then
\be
\lim_{n\to\infty}V_{4n+2}(\phi) = \lambda^2(1-\cos\phi) \equiv V_\text{sine-Gordon}(\phi),
\ee
which satisfies both $\min_\phi V(\phi) = 0$, $\alpha_{2i}>0$ for all $i$, and the coefficient of $\phi^{4n+2}$ is $\alpha_{4n+2} > 0$ as needed to ensure $V_{4n+2}(\phi) \to +\infty$ as $|\phi| \to \infty$. For these theories, the maximum number of degenerate minima is odd and, hence, there is a degenerate minimum at $\phi=0$, 
as in the sine-Gordon theory.

Lohe \cite{lohe} argued that both the $\phi^{4n}$ and $\phi^{4n+2}$ field theories limit onto the sine-Gordon theory with potential $\lambda^2(1-\cos\phi)$, while we showed that they limit onto field theories with potentials $ \lambda^2(1\pm\cos\phi)$, respectively. This is because Lohe \cite{lohe} only considered $\phi^{4n}$ field theories with a degenerate minimum at $\phi=0$, 
i.e., $V_{4n}(\phi) = \phi^2V_{4n-2}(\phi)$ (see \cite[Eq.~(10)]{lohe}), 
while this does not have to be the case in general [recall, e.g., the $\phi^{12}$ potential in \eqref{3.9}].
Nevertheless, the two limiting theories are indeed equivalent, as shown below.

\begin{figure}[h]
\includegraphics[width=\columnwidth]{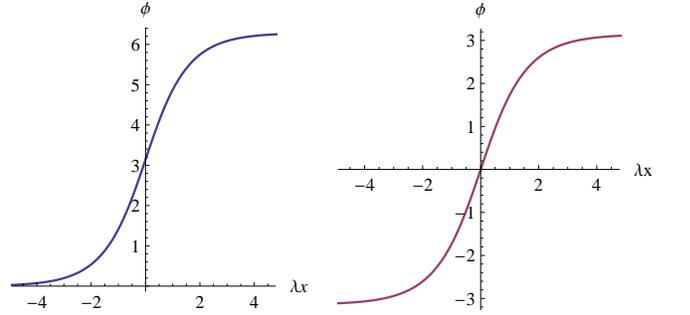}
\caption{(Color online.) Kink solutions \eqref{eq:limiting_kinks} of the $n\to\infty$ limiting field theories $V(\phi) = \lambda^2(1-\cos\phi)$ on the left (blue) and $V(\phi) = \lambda^2(1+\cos\phi)$ on the right (red).}
\label{fig:limiting_kinks}
\end{figure}

The limiting kink structures are easily found to be
\be\label{eq:limiting_kinks}
\tan(\phi/4) = 
\begin{cases}
\tanh(\lambda x/2), &V(\phi) = \lambda^2(1+\cos\phi),\\
e^{\lambda x}, & V(\phi) = \lambda^2(1-\cos\phi).
\end{cases}
\ee
Similarly, kink lattice solutions can also be obtained.
These two kinks are illustrated in Fig.~\ref{fig:limiting_kinks}.
The two limiting theories are equivalent through the transformation $\phi \mapsto \phi-\pi$, since $\cos(\phi-\pi) = -\cos\phi$ and $\tan(\phi/4-\pi/4) = [\tan(\phi/4) - 1]/[1 + \tan(\phi/4)]$ , which upon equating to $\tanh(\lambda x/2)$ and solving gives $\tan(\phi/4) = e^{\lambda x}$. It follows that both field theories are fully integrable.

We expect the corresponding statistical mechanics, correlation 
functions and PDFs of the $\phi^{4n+2}$ theories to approach, asymptotically as $n\rightarrow\infty$, 
those of the sine-Gordon theory derived in \cite{sutherland}.

\section{Conclusion} 

We have systematically studied high-order polynomial field theories 
(specifically, $\phi^8$, $\phi^{10}$ and $\phi^{12}$) describing
successions of phase transitions, and we obtained exact analytical (albeit implicit) 
kink solutions in different special cases of the theories with degenerate minima.
In view of \cite{footnote1},
steadily-translating kink solutions (with velocity $v$ and initial location $x_0$) 
can be obtained from the static kinks found herein through the Lorentz boost
\begin{equation}
\{x,t\}  \mapsto \left\{ \frac{x - x_0 - vt}{\sqrt{1-v^2}}, t \right\}.
\end{equation}
Similarly anti-kink solutions can be obtained through the transformation
\begin{equation}
\{x,\phi\} \mapsto -\{x,\phi\}.
\end{equation}

Some features of
the kink solutions found herein include asymmetry, power-law decay of their tails,
possibility of different kink types to have equal energy, and nonlinear phonons. 
The tail asymptotics that we derived for the kinks above could be used,
in conjunction with Manton's approach \cite{manton}, to compute
(asymptotically, for large separations) kink--kink and kink--anti-kink
interaction energies \cite{kks}. It would also be of interest to determine
whether the implicit kink solutions can be used to study interactions
via the collective-coordinate variational approximation techniques previously
applied to the $\phi^4$ \cite{sugiyama,csw}, $\phi^6$ \cite{gkl14} and sine-Gordon \cite{willis,cc} 
field theories.

The field theories considered above also possess pulse solutions confined 
to individual minima of the relevant potentials, however, such pulse solutions 
are beyond the scope of this work.  

Beyond meson physics \cite{ps,lohe}, the kink solutions obtained here 
correspond to domain walls in different ferroic materials such as ferroelectric 
and ferroelastic ones \cite{toledano,gl,mroz,pavlov,vc,piezo}.  It would 
be instructive to explore how asymmetric domain walls and nonlinear phonons 
affect the thermodynamic and physical properties of these materials.

\begin{acknowledgments}
A.K.\ acknowledges the hospitality of the Center for Nonlinear Studies 
and the Theoretical Division at LANL.  We gratefully acknowledge the support
of the U.S.\ Department of Energy through the LANL/LDRD Program for this
work, specifically a Richard P. Feynman fellowship to I.C.C.
LANL is operated by Los Alamos National Security, L.L.C.\
for the National Nuclear Security Administration of the 
U.S.\ Department of Energy under Contract No.\ DE-AC52-06NA25396.
\end{acknowledgments}


\begin{thebibliography}{99}

\bibitem{makhan} V. G. Makhankov, {\it Soliton Phenomenology} (Kluwer 
Academic Publ., Boston, 1990), Ch. VII. 

\bibitem{sanati} M. Sanati and A. Saxena, Am. J. Phys. {\bf 71}, 1005 
(2003). 

\bibitem{toledano} J.-C. Tol\'edano and P. Tol\'edano, {\it The Landau 
Theory of Phase Transitions} (World Scientific, Singapore, 1987); P. 
Tol\'edano and V. Dmitriev, {\it Reconstructive Phase Transitions} 
(World Scientific, Singapore, 1996).   

\bibitem{sonin} E. B. Sonin and A. K. Tagantsev, Ferroelectrics {\bf 98}, 
291 (1989). 

\bibitem{lohe} M. A. Lohe, Phys. Rev. D {\bf 20}, 3120 (1979). 

\bibitem{gufan} Y. M. Gufan, {\it Structural Phase Transitions} 
[in Russian] (Nauka, Moscow, 1982). 
 
\bibitem{gl} Y. M. Gufan and E. S. Larin, Dokl. Akad. Nauk SSSR {\bf 242}, 1311 (1978)
[Sovt. Phys. Dokl. {\bf 23}, 754 (1978)].
%Sovt. Phys. Dokl. {\bf 23}, 754 (1978); Izvest. Akad. Nauk. SSSR, Ser. Fiz. {\bf 43}, 1567 (1979). 

\bibitem{mroz} B. Mroz, J. A. Tuszynski, H. Kiefte, and M. J. Clouter,
J. Phys.: Condens. Matter {\bf 1}, 783 (1989).

\bibitem{pavlov} S. V. Pavlov and M. L. Akimov, Crystall. Rep. {\bf 44}, 
297 (1999). 

\bibitem{protein} A. A. Boulbitch, Phys. Rev. E {\bf 56}, 3395 (1997). 

\bibitem{vc} D. Vanderbilt and M. H. Cohen, Phys. Rev. B {\bf 63}, 094108
(2001); arXiv:cond-mat/0009337 [cond-mat.mtrl-sci].

\bibitem{piezo} I. A. Sergienko, Yu. M. Gufan, and S. Urazhdin, Phys. Rev. 
B {\bf 65}, 144104 (2002); arXiv:cond-mat/0109396 [cond-mat.mtrl-sci].

\bibitem{ms-ts} N. Manton and P. Sutcliffe, {\it Topological Solitons}
(Cambridge Univ. Press, Cambridge, UK, 2004), Ch. 5.

\bibitem{ps} T. H. R. Skyrme, Proc. R. Soc. A {\bf 262}, 233 (1961);
J. K. Perring and T. H. R. Skyrme, Nucl. Phys. {\bf 31}, 550 (1962).

\bibitem{Vach} T. Vachaspati, {\it Kinks and Domain Walls} (Cambridge
University Press, Cambridge, 2006).

\bibitem{zeld} Ya. B. Zeldovich, I. Yu. Kobzarev, and L. B. Okun, 
Zh. Eksp. Teor. Fiz. {\bf 67}, 3 (1974) [Sov. Phys. JETP {\bf 40}, 1 (1974)].

\bibitem{vm} D. Casta\~neda Valle and E. W. Mielke, 
Phys. Rev. D {\bf 89}, 043504 (2014).

\bibitem{jordan} P. M. Jordan, G. V. Norton, S. A. Chin-Bing, and A. Warn-Varnas,
Eur. J. Mech. B/Fluids {\bf 34}, 56 (2012).

\bibitem{GinzburgLandau} L. D. Landau, Zh. Eksp. Teor. Fiz. {\bf 7}, 19 (1937);
V. L. Ginzburg and L. D. Landau, Zh. Eksp. Teor. Fiz. {\bf 20}, 1064 (1950).

\bibitem{tinkham} M. Tinkham, {\it Introduction to Superconductivity} 
(McGraw-Hill, New York, 1996). 

\bibitem{boya} L. J. Boya and J. Casahorran, Ann. Phys. {\bf 196}, 361 
(1989). 

\bibitem{yang} J.-S. Yang and S.-Y. Lou, Z. Naturforsch. A {\bf 54}, 
195 (1999). 

\bibitem{casa} J. Casahorran, Phys. Lett. A {\bf 153}, 199 (1991). 

\bibitem{cooper} F. Cooper, L. M. Simmons Jr., and P. Sodano, Physica D 
{\bf 56}, 68 (1992). 

\bibitem{bazeia} D. Bazeia, M. A. Gonz\'alez Le\'on, L. Losano, and J. 
Mateos Guilarte, Phys. Rev. D {\bf 73}, 105008 (2006);
arXiv:hep-th/0605127 [hep-th].

\bibitem{bk} S. N. Behera and A. Khare, Pramana (J. of Phys.) {\bf 15}, 245
(1980).

\bibitem{bruce} A. D. Bruce, Adv. Phys. {\bf 29}, 111 (1980). 

\bibitem{leach} P. G. L. Leach, Physica D {\bf 17}, 331 (1985). 

\bibitem{sutherland} N. Gupta and B. Sutherland, Phys. Rev. A {\bf 14}, 
1790 (1976). 

\bibitem{BishopPhysD} A. R. Bishop, J. A. Krumhansl, and S. E. Trullinger,
Physica D {\bf 1}, 1 (1980). 

\bibitem{footnote1} Henceforth, by translation invariance 
of the equation of motion under  $x\mapsto x - x_0$, the kink has been centered
at $x = 0$ and, by mirror symmetry of the equation of motion under $x\mapsto -x$,
the ``$+$'' sign taken for the square root of $V$ above, without loss of generality,
for kink solutions \cite{ms-ts}.

\bibitem{sanati99} M. Sanati and A. Saxena, J. Phys. A: Math. Gen. {\bf 32}, 4311 (1999).

\bibitem{scal} D. J. Scalapino, M. Sears, and R. S. Ferell, Phys. Rev.
B {\bf 6}, 3409 (1972).

\bibitem{ks} J. A. Krumhansl and J. R. Schrieffer, Phys. Rev. B {\bf 11}, 3535 (1975).

\bibitem{manton} N. S. Manton, Nucl. Phys. B. {\bf 150}, 397 (1979).

\bibitem{kks} P. G. Kevrekidis, A. Khare, and A. Saxena, Phys. Rev. E {\bf 70}, 
057603 (2004); arXiv:nlin/0410045 [nlin.PS].

\bibitem{sugiyama} T. Sugiyama, Prog. Theor. Phys. {\bf 61}, 1550 (1979).

\bibitem{csw} D. K. Campbell, J. F. Schonfeld, and C. A. Wingate, Physica D {\bf 9}, 1 (1983).

\bibitem{gkl14} V. A. Gani, A. E. Kudryavtsev, and M. A. Lizunova, Phys. Rev. D {\bf 89}, 125009 (2014); arXiv:1402.5903 [hep-th].

\bibitem{willis} C. D. Ferguson and C. R. Willis, Physica D {\bf 119}, 283 (1998).

\bibitem{cc} I. Christov and C. I. Christov, Phys. Lett. A {\bf 372}, 841 (2008);
arXiv:nlin/0612005 [nlin.PS].

\end{thebibliography}
\end{document}